\documentclass{aa}
\usepackage{soul}
\usepackage{comment}
\usepackage{cuted}
\usepackage{hyperref}
\usepackage[normalem]{ulem}

\usepackage{comment}

\usepackage{xcolor}

\usepackage{graphicx}
\usepackage{float} 
\usepackage{subcaption}
\usepackage{txfonts}
\begin{document} 

   \title{Non-thermal emission from molecular clouds illuminated by continuous cosmic-ray accelerators}

   \subtitle{Application to the stellar cluster W43}

   \author{A. Inventar
          \inst{1}\fnmsep
          \thanks{\email{inventar@apc.in2p3.fr}}
          \and
          E. Peretti\inst{2,1}
          \and N. Tsuji \inst{4,5}     
          \and S. Gabici\inst{1,3}
          }

   \institute{Universit\'e Paris Cit\'e, CNRS, Astroparticule et Cosmologie, F-75013 Paris, France
         \and
             INAF Osservatorio Astrofisico di Arcetri Largo Enrico Fermi, 5, 50125, Firenze, Italy
         \and
             INAF – Osservatorio di Astrofisica e Scienza dello Spazio di Bologna, Via P. Gobetti 93/3, 40129 Bologna, Italy
        \and Institute for Cosmic Ray Research, The University of Tokyo, 5-1-5, Kashiwa-no-ha, Kashiwa, Chiba 277-8582, Japan
        \and Interdisciplinary Theoretical \& Mathematical Science Center (iTHEMS), RIKEN, 2-1, Hirosawa, Wako, Saitama 351-0198, Japan
             }

   \date{Received ?; accepted ?}

 
  \abstract
   {Cosmic rays escaping from their accelerators produce gamma-ray emission through interactions with ambient gas. While the scenario of an impulsive injection of cosmic rays has been extensively studied, that of a continuous injection in time has received comparatively less attention. 
   The latter scenario is particularly relevant for stellar clusters, which may contribute to a non-negligible fraction of Galactic cosmic rays.}
   {We intend to characterize the excesses of cosmic rays around their accelerators and to assess the detectability of the gamma-ray emission resulting from the interactions of cosmic rays in nearby molecular clouds. We focus in particular on star clusters and seek to apply these considerations to the case of W43.}
   {We derive analytical estimates for the cosmic ray excess around sources as a function of source properties and transport parameters. 
   We also evaluate under which conditions the predicted gamma-ray emission from illuminated molecular clouds can be detected by current and future instruments. 
   We then apply these results to explain the gamma-ray emission from the the W43 region.}
   {
   We find that producing detectable gamma-ray emission from molecular clouds illuminated by cosmic rays escaping from star clusters generally requires either suppressed diffusion relative to the Galactic value or anisotropic diffusion, with the latter scenario requiring a smaller global energetic. 
   We applied this approach to the case of W43 and by fitting its gamma-ray emission we were able to constrain the underlying CR transport parameters.
}
   {}

   \keywords{Cosmic rays -- acceleration of particles -- Stars: winds and outflows -- Gamma rays: stars -- ISM: clouds}
    \titlerunning{Non-thermal emission from molecular clouds illuminated by continuous CR accelerators}
   \maketitle

%

\section{Introduction}

Cosmic rays (CRs) escaping from their acceleration site diffuse in the surrounding medium and eventually mix with the CR sea that fills the entire Galaxy \citep{Gabici_SNR,marcowith2025}.
An excess of CRs has therefore to be expected within a given distance from the accelerator, and for a certain amount of time after CR escape.
If target material, most notably a molecular cloud (MC), is present in the vicinity of the accelerator, gamma-rays will be produced in interactions between CR nuclei and the dense gas.
Detecting these gamma-rays is important as it could reveal the presence of a CR accelerator in the region. 
After the pioneer paper by \citet{Atoyan_Aharonian}, most studies focused on the case of CR escaping from supernova remnant (SNR) shocks \citep[see e.g.][]{Gabici_Aharonian,Gabici2009,Fujita_SNR,Li_Chen,Gabici_W28,Torres_IC443,murase_2, Celli_SNR}. 
In this case, the injection of CRs is rather {\it impulsive}, i.e. limited to a short time (typically a small fraction of the SNR age).
The works mentioned above all considered an isotropic (3D) diffusion coefficient. 
Anisotropic (1D) diffusion of CRs escaping SNRs was discussed by, e.g., \citet{Nava}, and non-linearities (CR generation of magnetic turbulence) were studied by e.g. \citet{nava2016,dangelo2016,recchia2022,malkov2013,Ptuskin2008}.

The case of CRs escaping from systems where the injection is \textit{continuous} over an extended period of time comparable to the age of the accelerator has received much less attention, although it was originally considered in \citealt{Atoyan_Aharonian} and later applied in the context of microquasars \citep[e.g.][]{Bosch-Ramon2005}. 
This scenario is relevant for young massive stellar clusters, which have been proposed as potential accelerators of Galactic CRs \citep{cesmon1983,gabici2024,PeronICRC2025}. 
The acceleration of CRs in star clusters takes place at stellar wind termination shocks (WTS) \citep{Morlino,peron2024,peron2025}, and later, when cluster stars begin to explode, at SNR shocks \citep[e.g.][]{Vieu_superbubble}.
The combined energy input from these two sources remains roughly constants over several millions of years, and therefore star cluster can be considered as continuous injectors of CRs in the surrounding medium.

Here, we study the case of CR escape from continuous accelerators assuming both isotropic and anisotropic spatial diffusion of CRs in the ambient medium.
We also compare these results with the impulsive injection scenario and derive the gamma-ray emission expected from CR interactions with dense MCs, investigating under which conditions such emission can be detected by current and future gamma-ray observatories.

More specifically, Sec.~\ref{sec:2} presents analytic estimates for both continuous and impulsive CR injection scenario, under isotropic and anisotropic diffusion. These expressions can be readily used to estimate the CR excess around a given accelerator, as well as the maximum distance over which such an excess can be maintained. 
Moreover, we provide in Sec.~\ref{sec:CR_plots} a comprehensive characterization of the parameter space, including steady-state and time-dependent benchmark plots that can be applied to any system with known physical parameters. 
We further derive in Sec.~\ref{sec:gamm_clouds} the gamma-ray emission produced by MCs illuminated by escaping CRs, and provide benchmark results that can be easily applied to specific objects and directly compared with gamma-ray observations. 
Finally, in Sec.~\ref{sec:application}, we apply this formalism to the stellar cluster W43, investigating whether its observed gamma-ray emission can be explained by hadronic or leptonic interactions. 
We conclude in Sec.~\ref{sec:conclusion}.

\section{CR excess around sources: analytic estimates}
\label{sec:2}

In this Section we discuss both the case of isotropic and anisotropic diffusion of CRs escaping from a continuous source of particles.
For convenience, we will also briefly revisit the case of impulsive particle injection.

\subsection{Isotropic diffusion}
\label{sec:iso}

We consider first the case of an injection of CR protons which is impulsive in both space and time.
Let us assume that the injection takes place at $R = 0$, where $R$ indicates the distance from the CR accelerator (assumed to be point like), and at time $t = 0$\footnote{For simplicity, we neglect the fact that particles of different energy are expected to escape at different times \citep[see e.g.][]{Gabici2009}.}.
In the injection event, a total amount of energy $W_{CR}$ is released in form of CRs.
The differential energy distribution of CR particles is $N(E)$, where $E$ is the particle kinetic energy, so that
\begin{equation}
\label{eq:Wcr}
W_{CR} = \int_{E_{min}}^{E_{max}} {\rm d}E ~N(E) ~E
\end{equation} 
where $E_{min}$ and $E_{max}$ are the minimum and maximum energies of injected particles, respectively.
For definiteness, we limit ourselves to consider relativistic particles only ($E_{min} = m_p c^2$, $m_p$ is the proton mass) following a power law spectrum in energy:
\begin{equation}
N(E) = N_{min} \left( \frac{E}{E_{min}} \right)^{-\alpha} ~ .    
\end{equation}
When $\alpha$ is sufficiently larger than 2 and $E_{max} \gg E_{min}$, the expression above simplifies to:
\begin{equation}
\label{eq:inji}
N(E) \sim (\alpha - 2) \frac{W_{CR}}{\left( m_p c^2 \right)^2} \left( \frac{E}{m_p c^2} \right)^{-\alpha}
\end{equation}

Once injected, CRs undergo diffusive motion in the turbulent interstellar magnetic field.
The time evolution of the spatial distribution of CR particles around the accelerator can be obtained by solving the transport equation \citep{Ginzburg}, neglecting losses, as justified a posteriori in Sec.~\ref{sec:finite_size,adv,losses}:
\begin{equation}
\label{eq:transport}
\frac{\partial n(t,E,R)}{\partial t} = \frac{1}{R^2} \frac{\partial}{\partial R} \left( D(E) R^2 \frac{\partial n(t,E,R)}{\partial R} \right) + \frac{N(E)}{4 \pi R^2} \delta(t) \delta(R)
\end{equation}
where we assumed that diffusion is homogeneous and isotropic and characterized by an energy dependent spatial diffusion coefficient of the form $D(E) = D_0 [ E/(m_p c^2)]^{\delta}$.
The diffusion equation can be used when the characteristic particle diffusion length $l_d = \sqrt{6 ~D~ t}$ is smaller than $c t$. This condition can be rewritten as $t > t ~\mathrm{'} = 2 ~(D_0/10^{28} {\rm cm^2/s}) (E/m_p c^2)^{\delta}$~yr, which for very large particle energies (PeV domain), can give a quite large timescale.

Solving the transport equation gives the well known expression \citep{Atoyan_Aharonian}:
\begin{equation}
\label{eq:impulsive}
n(t,E,R) = \frac{N(E)}{\pi^{3/2} R_d(t,E)^3} e^{-\left( \frac{R}{R_d(t,E)} \right)^2}
\end{equation}
where $R_d(t,E) = \sqrt{4 D(E) t}$ is a characteristic length of the problem.
For $R \ll R_d$ the spatial distribution of CRs is nearly homogeneous, while for $R \gg R_d$ it drops quickly to zero.

As the goal of this Section is to provide order-of-magnitude estimates only, we do the crude approximation that the density of CRs is spatially uniform and equal to $n \approx N/[(4 \pi /3) R_d^3]$ within a distance $R_d$ from the accelerator, and vanishes elsewhere.
We also approximate the CR spectrum in the solar vicinity (Local Interstellar Spectrum, LIS) with a single power law $E^2 n_{LIS}(E) \approx 0.6 ~(E/m_p c^2)^{-0.7}$~eV/cm$^3$, which roughly reproduces the observed spectrum from the multi-GeV to the PeV domain (see \citealt{Recchia2024} and references therein for a compilation of data).

An excess of CRs above the typical Galactic level, characterized by $n\geq n_{LIS}$, is therefore expected for times smaller than:
\begin{equation} \label{eq:t_max_SNR}
\frac{\rm t_{max}}{\rm yr} \approx 2 \times 10^4 \left( \frac{\alpha - 2}{0.2} \right)^{\frac{2}{3}} \left( \frac{W_{CR}}{10^{50}{\rm erg}} \right)^{\frac{2}{3}} \left( \frac{10^{28}~{\rm cm^2/s}}{D_0} \right) \left( \frac{E}{m_p c^2} \right)^{1.8-\frac{2}{3} \alpha-\delta}
\end{equation}
after the injection time, and such an excess is found within a distance roughly equal to
\begin{equation} \label{eq:R_max_SNR}
R_d(t_{max}) \approx 60 \left(\frac{ \alpha - 2}{0.2} \right)^{\frac{1}{3}} \left( \frac{W_{CR}}{10^{50}{\rm erg}} \right)^{\frac{1}{3}} \left( \frac{E}{m_p c^2} \right)^{0.9-\frac{1}{3} \alpha} \rm pc
\end{equation}
from the accelerator.
The global energetic $W_{CR}$ has been normalized to 10\% of the typical supernova explosion energy, as expected if SNRs are the sources of CRs \citep[e.g.][]{gabici2019}.
For typical parameters such as $\alpha = 2.2$ and $\delta = 0.5$, one would get $t_{max} \sim 1.5 \times 10^4$~yr and $R_d \sim 90$~pc for $E = 10$~GeV, while $t_{max} \sim 4 \times 10^3$~yr and $R_d \sim 0.3$~kpc for $E = 10$~TeV, and $t_{max} \sim 2 \times 10^3$~yr with $R_d \sim 0.6$~kpc for $E = 1$~PeV.
The latter timescale is of the same order of $t\mathrm{'}$, which indicates that the diffusion approximation is valid for all but the largest (PeV) energies, and is valid for all particle energies of interest if the diffusion coefficient is significantly smaller than the reference value $D_0 \approx 10^{28}$~cm$^2$/s, $\delta = 0.5$. 

For times larger than $t_{max}$ or distances larger than $R_d(t_{max})$ the presence of the CR source will have no significant impact on the ambient CR intensity.
Finally, an excess of CR of an amount $\Delta = n/n_{LIS}$ will be found at a time $t \sim \Delta^{-2/3} t_{max}$ within a distance $R \sim \Delta^{-1/3} R_d(t_{max})$.
As an example, for the typical parameters listed above, an excess of an order of magnitude ($\Delta = 10$) of 10~GeV (10~TeV) CRs would be found at a time $\approx 3000$~yr ($\approx 900$~yr) within a distance $\approx 40$~pc ($\approx 140$~pc) from the SNR. 

The same reasoning can be applied to the case of a source releasing CRs continuously, starting from $t = 0$.
To model a stationary injection of CR at a rate of $\dot{W}_{CR}$, the last term on the right hand size of Eq.~\ref{eq:transport} should be substituted with $Q/(4 \pi R^2) \delta(R)$, where $Q$ is described by Eq.~\ref{eq:inji} with the substitutions $N \rightarrow Q$ and $W_{CR} \rightarrow \dot{W}_{CR}$.
Under these circumstances, the solution of the transport equation is:
\begin{equation}
\label{eq:continuous}
n(t,E,R) = \frac{Q(E)}{4 \pi D(E) R} \mathrm{erfc} \left( \frac{R}{R_d(t,E)} \right) ~ 
\end{equation}
where $\mathrm{erfc}$ is the complementary error function.

At this point, we can proceed as for the case of impulsive injection and approximate Eq.~\ref{eq:continuous} as $n \sim Q/4 \pi D R$ for $R < a \times R_d$ and 0 elsewhere.
To conserve the number of CR particles we set $a = 1/\sqrt{2}$.
Under these assumptions, an excess of CRs above the typical Galactic level can be found within a distance
\begin{equation} \label{eq:R_max_wind}
R_{max} \approx 6 \left(\frac{\alpha - 2}{0.2} \right) \left( \frac{\dot{W}_{CR}}{10^{37} {\rm erg/s}} \right) \left( \frac{10^{28}{\rm cm^2/s}}{D_0} \right) \left( \frac{E}{m_p c^2} \right)^{2.7-\alpha-\delta}  \rm pc
\end{equation}
from the accelerator.
Here, the CR power $\dot{W}_{CR}$ has been normalised to 10\% of the global mechanical power (stellar winds plus supernova explosions) of a rich stellar cluster \citep[e.g.][]{Celli2023,Inventar2026}.
If the same value of $D_0$ is adopted, the value of $R_{max}$ is much smaller than the typical values given by Eq.~\ref{eq:R_max_SNR} for the impulsive injection scenario.
Note that for $\alpha = 2.2$ and $\delta = 0.5$ the expression above becomes energy independent and for typical values of parameters reduces to $R_{max} \approx 6$~pc.

At a given distance $R < R_{max}$, the excess of CRs will be $\Delta \sim R_{max}/R$ for times larger than:
\begin{equation} \label{eq:t_min_wind}
\frac{t_{min}}{\rm yr} \approx \frac{200}{\Delta} \left(\frac{\alpha - 2}{0.2} \right)^2 \left( \frac{\dot{W}_{CR}}{10^{37} {\rm erg/s}} \right)^2 \left( \frac{10^{28}{\rm cm^2/s}}{D_0} \right)^3 \left( \frac{E}{m_p c^2} \right)^{5.4-2 \alpha-3 \delta} 
\end{equation}
which for $\alpha = 2.2$ and $\delta = 0.5$ and for typical values of the parameters gives $t_{min} \approx 6 ~ \Delta^{-1} ({\rm TeV}/E)^{0.5}$~yr.
This should be compared with the straight-line propagation time $t_* = R/c \sim 3~ (R/{\rm pc})$~yr to check whether or not the diffusion approximation is valid.
However, as both these time scales are always much shorter than the typical lifetime of a stellar cluster (which is of the order of millions of years) for any practical purpose we can consider the CR spatial distribution to be stationary in time and set $t_{min}$ equal to zero.
The straight-line propagation of particles will still impact on the spatial distribution of CRs, especially at the largest particle energies.
This is because particles injected at less than a time $t_*$ ago would follow a $n \propto R^{-2}$ spatial distribution, up to a distance $R_* \sim 3 D(E)/c \sim 0.3 (D_0/10^{28} {\rm cm}^2/{\rm s}) (E/m_p c^2)^{\delta}$~pc which, for sufficiently large energies, can be of the same order as $R_{max}$.
This implies that, for continuous injection and isotropic CR diffusion, the expressions above can be used only if the diffusion coefficient is suppressed with respect to the typical Galactic one.
As an example, fixing $\alpha = 2.2$, $\delta = 0.5$, $D_0 = 10^{27}$~cm$^2$/s and $\dot{W}_{CR} = 10^{37}$~erg/s, we get: $R_{max} \approx 60$~pc and $R_* \ll R_{max} $ for $E \ll 4$~PeV. 

\subsection{Anisotropic diffusion}
\label{sec:aniso}

Isotropic spatial diffusion of CRs is of course an idealised assumption. 
On spatial scales comparable or smaller than that of the magnetic field coherence length, CR will tend to propagate preferentially along the direction of magnetic field lines, and diffusion will be anisotropic \citep[e.g.][]{Ptuskin2008}.
Here we consider another idealised scenario where spatial diffusion is strictly one dimensional, i.e., particles propagate only along the direction $z$ defined by a perfectly aligned ambient magnetic field.
Any realistic mode of particle transport will be comprised between the two extreme idealised situations considered in the previous Section (3D and isotropic diffusion) and here (1D diffusion).

For an impulsive injection of CRs the one-dimensional transport equation is:
\begin{equation}
\label{eq:1D}
\frac{\partial n(t,E,z)}{\partial t} = \frac{\partial}{\partial z} \left( D(E) \frac{\partial n(t,E,z)}{\partial z} \right) + \frac{N(E)}{\pi R_s^2} \delta(t) \delta(z)
\end{equation}
where $R_s$ is the radius of the CR accelerator.
The solution of the equation is: 
\begin{equation} \label{eq:anisotropic_SNR}
n(t,E,z) = \frac{N(E)}{\pi^{3/2} R_s^2 R_d(t,E)} e^{-\left( \frac{z}{R_d(t,E)} \right)^2}
\end{equation}
that we will approximate as $n \approx N/[2 \pi R_s^2 R_d]$ for $z < R_d$ and 0 elsewhere. 

Before proceeding we should introduce another length scale, namely, the magnetic field coherence length $\lambda_B$, which defines the length of a magnetic flux tube as $L \sim \lambda_B/A^2$, where $A^2$ is the squared ratio between the random and total magnetic field \citep[see e.g.][]{Ptuskin2008}.
It follows that Eq.~\ref{eq:1D} is accurate only as long as $R_d \ll L$ or, equivalently, for times smaller than:
\begin{equation}
\label{eq:tstar}
t_{lim} =  7 \times 10^4 \left( \frac{L}{100~{\rm pc}} \right)^2 \left( \frac{D_0}{10^{28}~{cm^2/s}} \right)^{-1} \left( \frac{E}{m_p c^2} \right)^{-\delta} \rm yr
\end{equation}
which can be quite short for large particle energies.
If we limit the analysis to short time scales ($t \ll t_{lim}$), then the CR excess is:
\begin{equation} \label{eq:Excess_SNR_1D}
\Delta \approx 12 ~ \left( \frac{\alpha - 2}{0.2} \right) \left( \frac{W_{CR}}{10^{50} {\rm erg}} \right) \left( \frac{10~{\rm pc}}{R_s} \right)^{2} \left( \frac{100~{\rm pc}}{L} \right) \left( \frac{E}{m_p c^2} \right)^{2.7-\alpha} \left( \frac{t}{t_{lim}} \right)^{-\frac{1}{2}}
\end{equation}
where we normalized the shock radius to a typical SNR one.
The expression shows that much larger excesses can be obtained ($\Delta \gg 10$ for $t \ll t_{lim}$), with respect to the case of 3D diffusion. 

For times exceeding $t_{lim}$ particles will diffuse along field lines more than a coherence length.
The excess of CRs will drop quickly due to the wandering of magnetic field lines on scales larger than $L$ \citep[e.g.][]{Nava}. 
A full analytic prescription for the resolution of Eq.~\ref{eq:1D} with boundary conditions is given in Appendix.~\ref{sec:app_eq_imp}. 

The case of a continuous injection of CRs can be treated, as done in Sec.~\ref{sec:iso}, by changing the last term in Eq.~\ref{eq:1D} with $Q(E)/(\pi R_s^2) \times \delta (z)$.
If we assume that CRs diffuse mainly along magnetic field lines, and that they mix rapidly with background CRs after propagating a length $L$ (when the flux tube is disrupted), we can then solve Eq.~\ref{eq:1D} imposing the boundary condition $n(L) = 0$, to get a stationary solution:
\begin{equation} \label{eq:anisotropic_wind}
n(E,z) = \frac{Q(E)~L}{2 \pi R_s^2 D(E)} \left( 1 - \frac{z}{L}\right)    
\end{equation}
The overdensity of CRs around the injection region is then:
\begin{equation} \label{eq:Excess_wind_1D}
\resizebox{\hsize}{!}{$
\frac{\Delta}{10} \approx \left( \frac{\alpha - 2}{0.2} \right) \left( \frac{\dot{W}_{CR}}{10^{37}{\rm \frac{erg}{s}}} \right) \left( \frac{10 \, \rm pc}{R_s} \right)^{2} \left( \frac{10^{28}{\rm \frac{cm^2}{s}}}{D_0} \right)  \left( \frac{L-z}{50 \, \rm pc} \right) \left( \frac{E}{m_p c^2} \right)^{2.7-\alpha-\delta}
$}
\end{equation}
where we normalized the shock radius to a typical WTS radius \citep{weaver,gabici2024}. We note that this becomes energy independent for $\alpha = 2.2$ and $\delta = 0.5$.

Following the same reasoning as for the isotropic diffusion case we can check under which conditions the problem is almost stationary.
For standard parameters, $t_{lim}$ (Eq.~\ref{eq:tstar}) is significantly shorter than the age of a star cluster, and therefore the problem can be considered as stationary. 
However, if the diffusion coefficient is significantly smaller than the typical Galactic one, then time dependence may become important.
A full analytic prescription for the resolution of the transport equation in this case, considering time-dependency and boundary conditions, is given in Appendix.~\ref{sec:app_eq_cont}.

\subsection{Finite size of the accelerator, advection, and energy losses}
\label{sec:finite_size,adv,losses}

Before proceeding to a detailed parameter space exploration, we discuss several approximations that were adopted above. 

First, up to this point, the CR source has been treated as point-like. However, in more realistic situations, the acceleration region is spatially extended and particles may be injected over a finite volume. 
This can correspond to a shock surface, such as a stellar WTS or a FS of a SNR, or to a more extended region resulting from the combined action of several WTSs and SNs within a stellar cluster. 
If we retain the simpler shock picture and approximate the injection region by a thin shell of radius $R_s$, as defined in the previous subsection, the impact of this finite injection radius on the particle energy density can be assessed by replacing the term $\delta(R)$ in Eq.~\ref{eq:transport} (or $\delta(z)$ in Eq.~\ref{eq:1D}) by $\delta(R-R_s)$.

This modification mainly affects the solution at early times, i.e. when the diffusion length satisfies $R_{\rm diff} \lesssim R_s$. In this regime, the energy density at radii $R \gtrsim R_s$ is slightly enhanced with respect to the point-like case, since particles are injected closer to $R$ and therefore undergo less propagation. This effect progressively weakens as $R_{\rm diff}$ increases, and in the limit $R_{\rm diff} \gg R_s$ the solution converges to the point-source approximation. 
As $R_{\rm diff}$ increases with energy, the effect of finite size of the system is more relevant at low (GeV) energies, and can be often ignored at large ones (TeV or PeV).
A rigorous treatment accounting for a finite radius of an expanding SNR shock can be found in \citet{Li_Chen}. 
A similar qualitative behavior is expected in the anisotropic case, although small quantitative differences may arise owing to the different geometry.

As discussed above, for continuous injection the steady state solution applies in most cases. 
In this stationary regime, and in the 3D diffusion case, the solution for $R > R_s$ is identical to the point-like source limit. 
The same conclusion holds in the anisotropic scenario, with only minor geometrical corrections. Indeed, in Eq.~\ref{eq:anisotropic_wind}, the factor $L - z$ is simply replaced by $L + R_s - z$. 

As the changes are only minor in all these cases, we will keep considering point-like injection in the following. 

Additionally, in the specific case of stellar clusters, the presence of a wind-blown bubble beyond $R_s$ has been neglected so far. 
Within this structure, that can reach several tens of parsecs \citep{weaver,gabici2024}, a spatially varying transport regime is expected. 
Indeed, close to the shock (few parsecs), strong turbulence may significantly suppress diffusion potentially down to Bohm-like values \citep{Bykov2010}, while further away from the shock the diffusion coefficient can progressively increase as the level of magnetic turbulence decreases, and could approach the Galactic value towards the outer regions of the bubble.
Rather than relying on a specific phenomenological prescription for this spatial transition, we do not model the internal structure of the bubble explicitly. Instead, we will constrain the average effective diffusion coefficient within several tens of parsecs from the shock from observed gamma-ray emission, as discussed in Sec.~\ref{sec:gamm_clouds} and Sec.~\ref{sec:application}, and keep the Galactic diffusion coefficient as a reference scale.

\begin{figure}
   \centering
   \includegraphics[width=0.9\columnwidth]{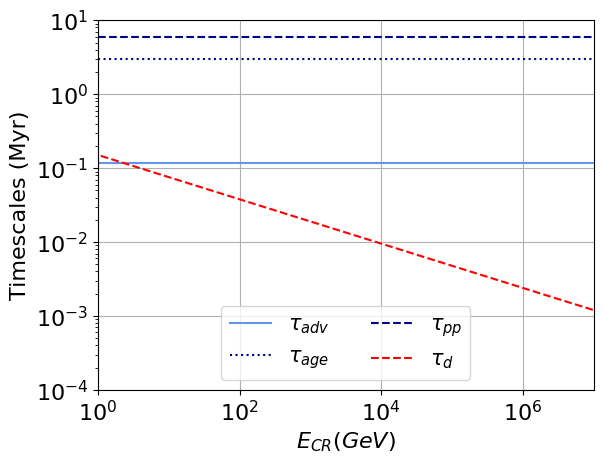}
   \caption{Timescales of the problem: age of the cluster, proton energy loss time, advection and diffusion timescales for the parameters given in the main text.}
    \label{fig:timescales_hadronic}%
 \end{figure}

Moreover, within the bubble, advection may also modify the transport equation. Following \citet{Morlino}, keeping the bubble at a constant mass density implies a downstream velocity profile $u(R)=u_2(R_s/R)^2$, where $u_2$ is the downstream wind speed ($=u_w/4$ for a strong shock, noting $u_w$ the upstream wind speed). This leads to a characteristic advection timescale at a distance $R$ from the center $\tau_{\rm adv}(R)= \int_{R_s}^{R} dR/u(R)\simeq R^3/(3R_s^2 u_2)$. This is compared with the 3D diffusion timescale $\tau_{d}(R,E)=R^2/(6D(E))$ in Fig.~\ref{fig:timescales_hadronic} for $R=30~\rm pc$, $R_s=10~\rm pc$, $u_w=3000~\rm km/s$, and $D(E)=3\times 10^{26}({E}/{\rm GeV})^{0.3}~\rm cm^2/s$. 
Even with such a small diffusion coefficient and large wind velocity, advection remains negligible over the entire energy range considered, including low energies. It would only become important closer to the shock or for even smaller diffusion coefficients, with the corresponding regime discussed in \citet{Morlino}.
Therefore, these effects of the wind bubble are neglected in the following.

\begin{figure*}
   \centering
   \includegraphics[width=0.99\textwidth]{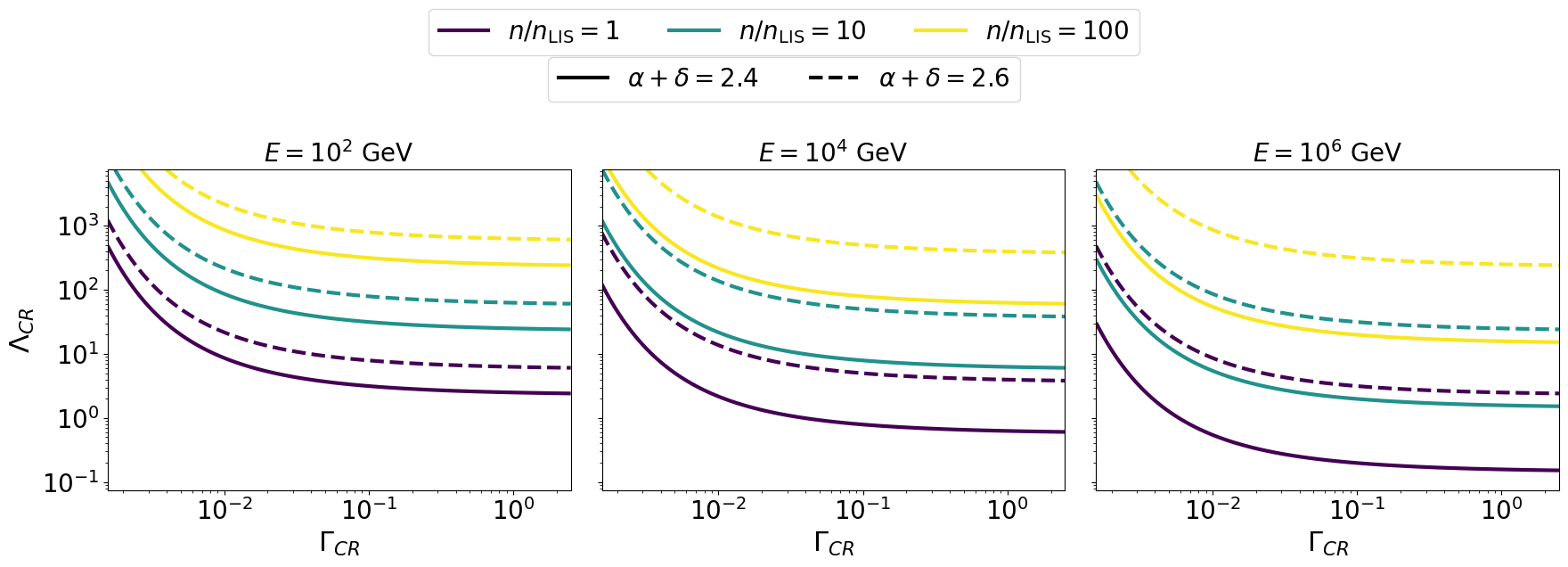}
   \caption{Values of $\Lambda_{CR}$ and $\Gamma_{CR}$ (Eq.~\ref{eq:Lambda_CR} and \ref{eq:Gamma_CR}) corresponding to an excess of CRs over the Galactic background equal to 1, 10, and 100 (violet, green, and yellow curves, respectively). The three panels corresponds to different particle energies, and solid and dashed lines refer to different values of $\alpha+\delta$.}
    \label{fig:adim_par_wind}%
 \end{figure*}

Finally, energy losses due to nuclear interactions with the ambient gas have also been neglected above. The corresponding energy loss time $\tau_{pp}\sim 6\times 10^7 (\rm cm^{-3}/n_0)$~yr is shown in Fig.~\ref{fig:timescales_hadronic} for a medium of density $n_0=10~\rm cm^{-3}$, together with a young star cluster age of $\sim 3~\rm Myr$. Although the density inside a wind bubble is expected to be much lower \citep{Gupta2018}, such larger values may occur in the presence of dense molecular material \citep{blasi2025}. This timescale remains however significantly longer than the other relevant ones, and losses become important only when $t \gg \tau_{pp}$. In this regime, the flux is strongly suppressed (as it can be seen in the impulsive case by the $1/R_d^3$ dependence), such that the impact of losses will be neglected in the following.

\section{CR excess around sources: varying parameters}
\label{sec:CR_plots}

In the previous Section we derived the main scalings of continuous and impulsive injection scenarios, both for isotropic and anisotropic transport, and highlighted the dependence of the CR excess on the physical parameters of the system. 
To complement these analytical considerations, we now perform  an exploration of the parameter space. 
We consider here the continuous injection case only, and we refer the reader to Appendix~\ref{app:impulsive} for a discussion of the impulsive injection scenario.

In the case of continuous injection and three-dimensional diffusion (ignoring the ballistic regime), the CR excess at a given energy is entirely determined by the three quantities $\Lambda_{CR}$, $\Gamma_{CR}$, and $s$ that allow us to rewrite Eq.~\ref{eq:continuous} as 
\begin{equation}
    n \approx 4 \times 10^{-8} ~\Lambda_{CR}~ (E/\rm GeV)^{-s} ~\mathrm{erfc}(0.1/\Gamma^{1/2}_{CR}) ~ \rm erg^{-1} cm^{-3} 
\end{equation}
where
\begin{eqnarray}
\label{eq:Lambda_CR}
    \Lambda_{CR} &=& \frac{\alpha-2}{0.2}
~\frac{\dot{W}_{CR}}{10^{37}\,\rm erg/s}
~\frac{10^{28}\,\rm cm^2/s}{D_0}
~\frac{50\,\rm pc}{R} \\
\label{eq:Gamma_CR}
    \Gamma_{CR} &=& \frac{D_0}{10^{28}\,\rm cm^2/s}
~ \left ( \frac{E}{\rm GeV}\right)^{\delta}
~\frac{t}{3\,\rm Myr}
~\left(\frac{50\,\rm pc}{R}\right)^2 \\
s &=& \alpha+\delta 
\end{eqnarray}

The values of $\Lambda_{CR}$ and $\Gamma_{CR}$ that result in an excess of CRs above the Galactic sea of $n/n_{LIS} =$~1, 10, and 100 are plotted in Fig.~\ref{fig:adim_par_wind}. 
The three panels refer to different particle energies, and solid and dashed lines refer to two different values of $\alpha+\delta$.

Notice that the parameter $\Gamma_{CR} \propto t$ can be interpreted as a normalized time, i.e., as a proxy for the temporal evolution of the system. 
Fig.~\ref{fig:adim_par_wind} shows that as time increases the normalization required to produce a given CR excess decreases until it reaches a plateau corresponding to the steady-state regime, which occurs when $\Gamma_{CR} \gtrsim 10^{-2}$. 
Moreover, the value of $\Lambda_{CR}$ required to produce a given excess increases with $\alpha+\delta$, while it decreases with particle energy. 
A particularly important result of Fig.~\ref{fig:adim_par_wind} is that, for typical ages of stellar clusters of a few Myr or more, the system will be in steady state, unless quite extreme parameters are adopted. 
This can be seen from Eq.~\ref{eq:Gamma_CR}, showing that in order to have $\Gamma_{CR} < 10^{-2}$ one would need to assume extremely small values for $D_0$ compared to the ISM one, or unreasonably large distances from the accelerator (or a combination of the two).

Finally, assuming to be in steady state and considering the case $s=2.4$, the condition to have an excess ($n/n_{LIS} > 1$) at 10$^2$, 10$^4$, and 10$^6$~GeV is respectively $\Lambda_{CR} \gtrsim 3$, $\Lambda_{CR} \gtrsim 0.6$ and $\Lambda_{CR} \gtrsim 0.2$ .
As $\Lambda_{CR} = 1$ corresponds to a standard choice of parameters, we can conclude that having an excess of CRs within tens of parsecs from the source is quite likely, but a large excess would require a small value of the diffusion coefficient.

A representation of the parameter space as the one shown in Fig.~\ref{fig:adim_par_wind} is not possible for the 1D case, because the expression of $n$ involves several coupled terms (see Eq.~\ref{eq:1D_wind_time} in Appendix~\ref{sec:app_eq_cont}).
Nevertheless, in the steady-state regime a comparison with the isotropic diffusion case is still possible.
In this case Eq.~\ref{eq:anisotropic_wind} can be rewritten as 
\begin{equation}
    n \approx 2\times 10^{-6}~ \Lambda_{CR,~1D} ~(E/\rm GeV)^{-s_{1D}} ~ \rm erg^{-1}cm^{-3}
\end{equation}
where
\begin{align}
\label{eq:Lambda_CR_1D}
\Lambda_{CR,~1D} &= \frac{\alpha-2}{0.2}
\, \frac{\dot{W}_{CR}}{10^{37}\,\rm erg/s}
\, \frac{10^{28}\,\rm cm^2/s}{D_0}
\,  \left(\frac{10\,\rm pc}{R_s}\right)^2
\,  \left(\frac{L-z}{50~\rm pc}\right)\\
s_{1D} &= \alpha+\delta 
\end{align}
leading to a CR flux $\sim 50$ times larger than in the isotropic case, if the same value for the diffusion coefficient and CR power are adopted.
As discussed above, it is easier to obtain large excesses of CRs above the Galactic background if particles leaving the source are confined within a magnetic flux tube, and are not free to disperse isotropically.


\section{Gamma-rays from molecular clouds}
\label{sec:gamm_clouds}

\begin{figure*}
   \centering
   \includegraphics[width=0.9\textwidth]{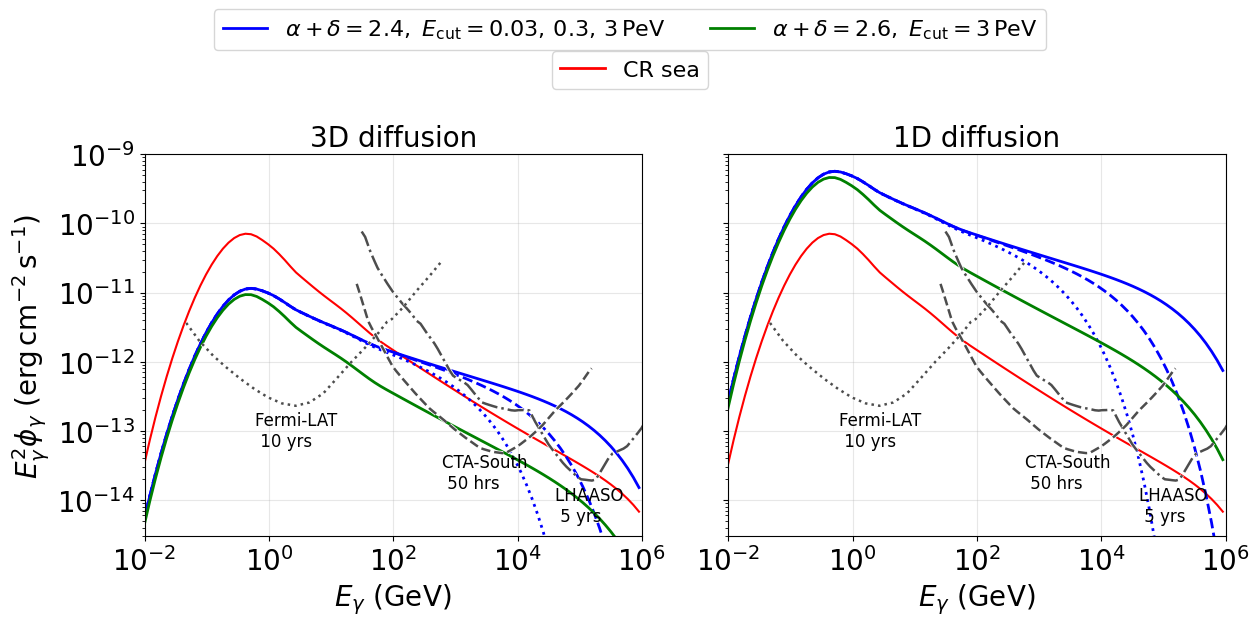}
   \caption{Gamma-ray
   emission from a MC ($A_5 = 1$)
   illuminated by CRs coming from a continuous source ($\Lambda_{CR} = \Lambda_{CR,1D} = 1$, $\Gamma_{CR} \gg 10^{-2}$). The left (right) panel refers to 3D (1D) diffusion. 
   Different colors (linestyles) refer to different spectral slopes (cutoff energies).
   The gamma-ray from interactions of Galactic CRs is shown in red. Sensitivities of gamma-ray instruments are also shown.}
   \label{fig:gamma_wind_steady}%
 \end{figure*}

Having identified which regions of the parameter space can produce significant CR excesses, we now investigate the corresponding gamma-ray emission generated through hadronic interactions.
Indeed, escaping accelerated protons as well as protons in the Galactic CR sea can interact with target protons in dense environments and produce gamma-rays through the reaction $p+p \rightarrow p+p+\pi^0$, followed by the decay $\pi^0 \rightarrow 2\gamma$. 
Here, we consider a spherical MC of mass $M$, located at a distance $R$ from the accelerator.
The cloud volume and radius are $V$ and $R_c$, respectively, and $d$ is the distance from the Sun.
The resulting gamma-ray flux can be written as:
\begin{equation}
\label{eq:flux_gamma}
    \phi_{\gamma}=\frac{c}{4\pi d^2}\frac{M}{m_p}
    \int \frac{d\sigma}{dE_\gamma}(E,E_{\gamma})\,n(E)\,dE
\end{equation}
where the differential cross section is taken from \citet{Kafexhiu} and $n$ is the differential CR density (escaping CRs plus CRs in the Galaxtic sea). 

In fact, the observed gamma-ray emission also includes the contribution from the hadronic interactions of Galactic CRs with the interstellar gas along the entire line of sight.
To estimate the importance of this contribution, the column density of the cloud 
$N_{col}^{cloud} \sim (M/m_p V) \times (2R_c) \sim 7 \times 10^{22} (M/10^5~\rm M_\odot) (10~\rm pc/R_c)^2 ~\rm cm^{-2}$ has to be compared with the typical gas column density along the Galactic disk $N_{col}^{gas}=n_{ISM} \times L \sim 3 \times 10^{22} (n_{ISM}/{\rm cm}^{-3}) (L/10~{\rm kpc}) ~\rm cm^{-2}$, where $n_{ISM}$ and $L$ are the typical interstellar medium density and pathlength.
This shows that for giant MCs, as those we consider, of mass exceeding $10^5 M_{\odot}$ this contribution can be often neglected, while for less massive and/or more extended clouds it can not 
\citep[see e.g.][]{Aharonian2020}.
In the following, we will neglect this contribution, and we also assume that CRs can freely penetrate MCs, an assumption that is quite reasonable especially at large particle energies.
For simplicity, in order to compute the gamma-ray emission from a MC, we will take the spectrum of CRs in its centre as representative for the entire cloud volume.

\subsection{Benchmark spectra in steady state}
\label{sec:plot_gamma_1}

We first focus on the configuration where the MC is illuminated by a source injecting CRs continuously and at a stationary rate\footnote{The results for an impulsive source are shown in Appendix~\ref{app:impulsive}.}.
We will consider both 3D (Eq.~\ref{eq:continuous}) and 1D diffusion (Eq.~\ref{eq:anisotropic_wind}).
For sufficiently large times ($\Gamma_{CR} \gg 10^{-2}$), the emission is entirely characterized by only two quantities: a normalization factor and the total spectral slope $\alpha+\delta$.
The normalisation factor is:
\begin{eqnarray}
\label{eq:Lambda_gamma}
    \Lambda_{\gamma,3D} &=& \Lambda_{CR}~\rm A_5 \\
\label{eq:Lambda_gamma_2}
    \Lambda_{\gamma,1D} &=& \Lambda_{CR,~1D}~\rm A_5
\end{eqnarray}
where $A_5=(M/10^5M_{\odot}) \left({\rm kpc}/d\right)^2$ is the mass of the MC over its distance squared normalized to reference values.

Fig.~\ref{fig:gamma_wind_steady} shows the gamma-ray emission from a benchmark system characterized by $A_5 = 1$ and $\Lambda_{\gamma,3D} = 1$ (left) or $\Lambda_{\gamma,1D} = 1$ (right panel).
We include an exponential cutoff in the CR proton spectrum to account for the finite maximum energy achievable in the accelerator (dotted, dashed, and solid lines refer to maximum energies of 0.03, 0.3, and 3 PeV, respectively).

Blue and green lines show the contribution to the gamma-ray from the cloud resulting from interactions of CRs that escaped the acceleration adopting a different choice for the total spectral slope ($\alpha + \delta = 2.4$ and 2.6).
As discussed above, a larger gamma-ray flux (about 50 times, in this case) is expected for the 1D setup.
Notice that spectra can be readily rescaled to any specific source-cloud configuration by simply changing the normalization parameter (as we assume stationarity, the only effect would be to shift spectra up or down).
The total gamma-ray emission from the cloud is the sum of the contribution from CR escaping from the source and that of the sea of Galactic CRs.
The latter contribution is shown in red in the figure.

As anticipated from Fig.~\ref{fig:adim_par_wind}, for isotropic diffusion and for total spectral slopes larger than $\alpha + \delta \sim 2.5$, the CR sea contribution dominates over the source one at all energies.
An increase in the normalization would require either augmenting the value of $A_5$ (more massive and/or closer cloud) and/or that of $\Lambda_{CR}$ (more powerful source and/or smaller diffusion coefficient and/or smaller distance between source and cloud).
For harder spectra ($\alpha + \delta \lesssim 2.5$), the source contribution exceeds the CR sea at high energies ($E_{\gamma} \gtrsim 100~\mathrm{GeV}$) in the benchmark case.
On the other hand, in the anisotropic diffusion case the source dominates over the CR sea at all energies in the benchmark case.

The point-source differential sensitivities of several gamma-ray instruments spanning the GeV--PeV range (Fermi-LAT, CTA-South and LHAASO, \citealt{Addazi2022}) are also shown in Fig.~\ref{fig:gamma_wind_steady}. 
For the benchmark cases considered in the figure, the predicted emission is above the sensitivity for all three instruments. 
However, these sensitivities correspond to point-like sources, while clouds are often extended. 
If the extension of the cloud is $\theta_{\rm cl}$, then the instrument sensitivity is degraded by a factor roughly of the order of
$(\theta_{\rm cl}^2+\theta_{\rm PSF}^2)^{1/2}/\theta_{\rm PSF}$,
where $\theta_{\rm PSF}$ is the instrument point-spread function. 
Since giant MCs can reach angular sizes of the order of $\sim 1^\circ$, the chances of detectability may be substantially reduced, particularly for IACTs such as CTA, whose PSF is quite small (see e.g. \citealt{Celli2024} for a detailed discussion). 

\subsection{Time dependent gamma-ray emission }
\label{sec:plot_gamma_2}

Finally, we investigate the evolution of the gamma-ray spectra in the regime where a steady state has not been established, which is characterised by small value of the parameter $\Gamma_{CR}$.
As done above, we consider both the cases of 3D (Eq.~\ref{eq:continuous}) and 1D diffusion (Eq.~\ref{eq:1D_wind_time}) for a continuous source of CRs\footnote{The impulsive source case is treated in Appendix~\ref{app:impulsive}.}.
As discussed in Sec.~\ref{sec:CR_plots}, such a regime is expected only under rather extreme conditions, namely young source ages, strongly suppressed diffusion coefficients, and/or large source-cloud separations.

\begin{figure}
   \centering
   \includegraphics[width=0.45\textwidth]{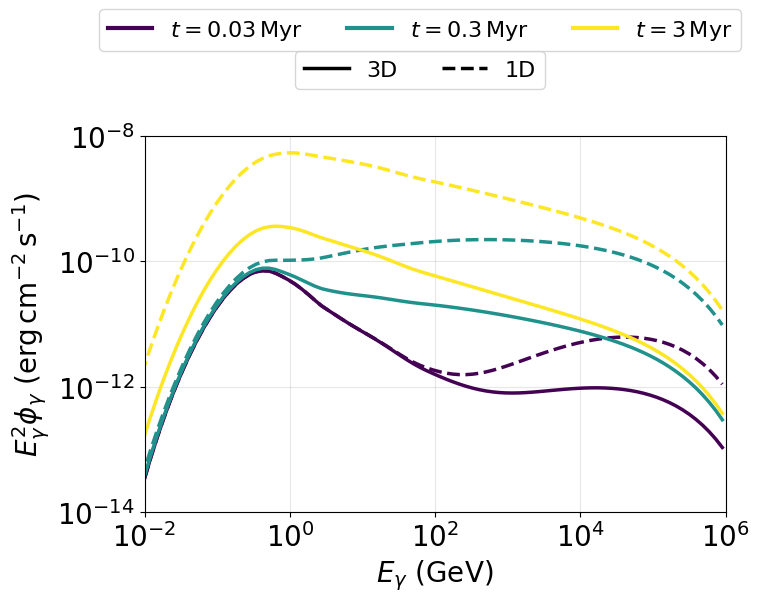}
   \caption{Gamma-ray emission from a MC illuminated by CRs coming from a continuous source. Solid (dashed) lines refer to 3D (1D) diffusion, and different colors represent different times. An exponential cutoff is present in the CR spectrum at an energy $E=3 \rm~PeV$.
   Other parameters are fixed as: $\dot{W}_{CR}\rm A_5 =10^{37} \rm ~erg/s$, $\rm R=50~pc$, $L= 100~\rm pc$, $D_0=10^{26} \rm cm^2/s$, $\delta=0.3$, $\alpha=2.2$.}
    \label{fig:gamma_wind_time}%
 \end{figure}

To illustrate this behavior, we first examine the effect of varying the source age while keeping all other parameters fixed to values for which the steady-state regime is not attained. Fig.~\ref{fig:gamma_wind_time} shows the resulting gamma-ray spectra (the sum of the contributions coming from interactions of both the CRs escaped from the source and those in the CR sea) for ages ranging from $0.03$ Myr to $3$ Myr.
Solid (dashed) curves refer to 3D (1D) diffusion, and have been computed for a quite small value for the diffusion coefficient ($D_0 = 10^{26}$~cm$^2$/s), and fixing the other relevant parameters as indicated in the figure caption\footnote{See Appendix \ref{app:nonsteady} for a more complete study of parameter space, where also the source-cloud distance $R$, the diffusion coefficient normalization $D_0$ and its slope $\delta$ for the energy dependence are varied.}.

In both the 3D and 1D cases, a similar non-trivial energy dependence emerges at early times: the low-energy part of the spectrum is strongly suppressed whereas high-energy particles can reach the cloud more efficiently, leading to a significant deviation from a simple power-law shape. 
This is consistent with the results of \citet{Merten2026}, where advection was also taken into account.

\section{Application to the star cluster W43-main and its surroundings}
\label{sec:application}

The modeling described above can be used to interpret non-thermal emission in the vicinity of stellar clusters and to constrain key physical parameters, such as the CR acceleration efficiency or the CR diffusion coefficient. 
Detailed studies have already been carried out for a number of star clusters and associations exhibiting GeV–TeV gamma-ray emission, including Westerlund 1 \citep{harer2023,Lemoine_2025}, Westerlund 2 \citep{aharon2007,Mestre2021}, and Cygnus OB2 \citep{LHAASO_cygnus, harer2025}. Other systems have been analyzed primarily in the GeV band, such as NGC 3603 \citep{Saha2020,Rocamora2025}, NGC~6611 \citep{peron2025}, and RCW~38 \citep{peron2024}. However, the recent detection of GeV–TeV emission in the vicinity of the W43-main cluster has received comparatively limited attention \citep{Yang,Wang}. In this context, we provide a complementary description and interpretation of this object.

\subsection{Properties of the W43 region and of the massive star cluster in W43-main}

W43 (=G30.8–0.2) lies on the Scutum–Crux spiral arm and is one of the main sites of massive star formation in the inner Galaxy \citep{Smith}.
The region hosts two of the largest MC structures in the first Galactic quadrant: W43-Main and W43-South. 
W43-Main is an extended $H_{\rm II}$ region powered by a compact stellar cluster containing OB and Wolf–Rayet (WR) stars. Three bright stars are detected in this cluster: a WR of WN7 subtype (WR 121a, \citealt{roscro2015}) that is suspected to be in a binary system, as well as two luminous O-type stars \citep{Blum}. These three objects are also known as W43-1, W43-2, and W43-3 respectively. 

\citealt{Smith} inferred that W43-Main releases a Lyman continuum luminosity of $10^{51}$ ionizing photons per second. In addition, its far-infrared luminosity, produced by dust re-emission, has been estimated at $3.5 \times 10^6 L_{\odot}$ \citep{Lester1985, Bally}, or about $1.5\times 10^{40}$ erg/s.
When expressed in terms of stellar equivalents, this power corresponds to roughly one hundred O7-type stars \citep{Vacca1994}. As remarked by \citet{Blum}, such values are comparable to those of the cluster NGC 3603 and only a few times smaller than the luminosity of R136 in the Large Magellanic Cloud, both of which are detected in gamma-rays \citep{Rocamora2025,Aharonian2024}. This suggests that the star cluster powering W43-main hosts a substantial population of massive stars that remain undetected.

Due to the intense star formation occurring in its central region, W43-Main is often described as a “mini-starburst”. Four very massive and dense cores  (W43-MM1 to W43-MM4) were initially proposed as potential sites of high-mass star formation by \citealt{Motte}. Subsequent Herschel observations, covering far-infrared to sub-millimeter wavelengths, confirmed this scenario and revealed a complex arrangement of filaments and chimneys within the region \citep{Bally}. 
The star formation activity in W43 is very high, at the level of $0.01 M_{\odot}$/yr , and might increase by an order of magnitude in the future \citep{Nguyen_2011_w43, Louvet_w43_herschel}. 

This complex thus stands out as an exceptionally massive structure. Detailed studies of molecular and atomic cloud tracers \citep{Nguyen_2011_w43,Motte2} indicate that the total gas mass approaches $10^7 M_{\odot}$, with dense clumps making up roughly 10 $\%$ of this mass, and the remainder residing in larger clouds. 
Such a significant accumulation of matter establishes a strong gravitational potential, which may itself draw in nearby GMCs \citep{Motte2}. 

The distance of this complex is estimated at approximately 5.5 kpc \citep{Zhang}, while the age of the central massive stellar cluster is estimated to lie between 5 and 10 Myr \citep{Bally}. In the following, we adopt a fiducial value of $5~\rm Myr$, as the presence of a WR star suggests that the cluster is likely located toward the younger end of this interval.

\subsection{Mechanical Luminosity}
\label{sec:mech_lum}

\begin{figure}
    \centering    \includegraphics[width=0.9\columnwidth]{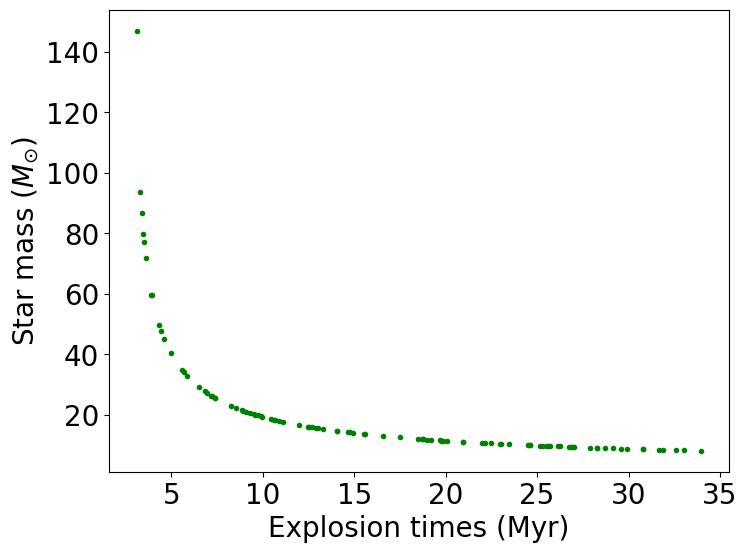}
    \caption{Explosion times and stellar masses for a simulated cluster containing $N=100$ O stars.}
    \label{fig:mass_distribution}
\end{figure}

The observed IR luminosity ($L=1.5\times 10^{40}$ erg/s) provides a lower limit on the bolometric luminosity of the cluster, since a large fraction of the UV/optical radiation emitted by massive stars is absorbed by dust and re-emitted in the infrared. We therefore adopt a conservative bolometric luminosity of $2\times 10^{40}~\rm erg/s$ in the following.
This value can be used to estimate the population of massive stars in the cluster, and consequently the associated wind and supernovae mechanical luminosities $L_w$ and $L_{SN}$. 
Denoting the CR acceleration efficiencies at the WTS and SNR shocks as $\eta_{w}$ and $\eta_{SN}$, respectively, we set $\dot{W}_{CR}=\eta_{w} L_w + \eta_{SN}L_{SN}$. 
We model the time evolution of a stellar cluster population composed of an initial number $N$ of massive O stars (which may later evolve through a WR phase), using a Salpeter initial mass function \citep{Salpeter1955} and following the approach of \citet{Vieu_superbubble}. The stellar lifetimes as a function of mass are taken from \citet{Limongi_chieffi} (see Fig.~\ref{fig:mass_distribution}).

At each timestep, the bolometric luminosity of the cluster is computed by summing the contributions of all individual stars. 
For progenitor masses larger than $25\,M_{\odot}$, a WR phase of duration of $t=3.2\times 10^5~\rm yr$ is added after the main sequence phase.
The mass–luminosity relations for O and WR stars are taken from \citet{Menchiari2024} and \citet{crowth2007}, respectively.

\begin{figure}
    \centering
    \includegraphics[width=0.9\columnwidth]{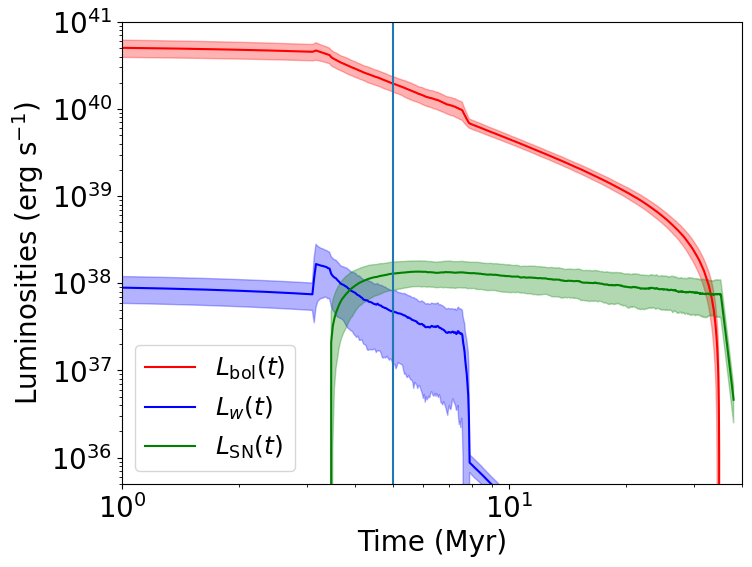}
    \caption{Evolution of the bolometric luminosity (red), mechanical wind luminosity (blue), and SNRs mechanical luminosity (green) of a star cluster with $N=100$ O stars. Lines show the mean values, while shaded bands indicate the standard deviations over the Monte Carlo realizations. The vertical line marks $5~\rm Myr$, the estimated age of the cluster.}
    \label{fig:luminosities}
\end{figure}

We then infer the mean initial number of stars $N_*$ by requiring that the ensemble-averaged bolometric luminosity after 5 Myr, obtained from 1000 Monte-Carlo realizations, matches the observed value. 
This procedure yields to $N_* \simeq 100$ initial massive stars, consistent with previous estimates \citep{Vacca1994}. 
Fig.~\ref{fig:mass_distribution} shows a representative Monte-Carlo realization of a simulated cluster, while Fig.~\ref{fig:luminosities} displays the corresponding ensemble-averaged bolometric luminosity and its standard deviation, together with the mechanical powers of stellar winds and SNRs.

The mechanical wind luminosity of the cluster has been computed at each timestep by summing the individual stellar contributions using the prescriptions by \citet{Seo}, applied to O stars and to WR stars in the WN or WC phases. WN and WC phases are assumed to occur when the mass of the star is respectively larger or smaller than $15\,M_{\odot}$  \citep{crowth2007}. 
The obtained averaged mechanical luminosity at 5 Myr, is $L_{w}= (4.8 \pm 3.6)\times 10^{37}\,\rm erg/s$. 
The time evolution of the wind mechanical luminosity and its dispersion are shown in Fig.~\ref{fig:luminosities}.

In addition, we estimate the mechanical power injected by supernova explosions within the cluster. Following \citet{Vieu_superbubble}, we assume a turbulence relaxation time of $1~\rm Myr$ for each SNR, with an exponential decay thereafter. The resulting average power is also shown in Fig.~\ref{fig:luminosities}, yielding to $L_{\rm SN}=(1.3 \pm 0.5)\times 10^{38}\,\rm erg/s$ at 5~Myr.
The two contributions (winds and supernovae) are of the same order of magnitude, and the acceleration efficiencies are also expected to be comparable \citep{peron2024}. We can thus write $\dot{W}_{CR}\sim \eta L_m$ with $\eta$ the average acceleration efficiency and $L_m$ the total mechanical power of the cluster that can be safely approximated as $\sim 10^{38}\,\rm erg/s$, which is adopted in the following.

For comparison, previous estimates of the stellar cluster's mechanical luminosity were slightly lower, with values of $10^{37}\,\rm erg/s$ in \citet{Wang} and $3\times 10^{37}\,\rm erg/s$ in \citet{Yang}. 
This is because their estimates were based on only three detected bright stars, potentially leading to an underestimate of the total stellar wind power, and did not account for the cumulative contribution of successive supernovae.

\subsection{Gamma-ray and X-ray/radio observations}

Observations of this region have been performed across different wavelengths to probe non-thermal emission potentially associated with CRs.
At very-high energies, \citet{Chaves2008} first reported the detection of a gamma-ray source in this region using H.E.S.S. observations. The source was subsequently studied within the H.E.S.S. Galactic Plane Survey \citep{HESS} and more recently detected by LHAASO, as reported in the First Catalog paper \citep{LHAASO_catalog} as well as in a dedicated analysis \citep{Wang}. 
These gamma-ray sources are HESS J1848-018 and LHAASO J1848-0153u, respectively.
The centroids of the sources are separated by only $\sim 0.1^\circ$, HESS J1848-018 being best fitted by a 2D Gaussian with 1$\sigma$ extension of $\sim 0.25^\circ$, while LHAASO J1848-0153u is modeled with an ellipse of semi-major axis $\sim 0.6^\circ$ \citep{Wang}, suggesting that they likely trace the same underlying emission region. However, in both cases the centroid is offset by $\sim 0.2^\circ$ with respect to the position of the stellar cluster.
At GeV energies, a dedicated analysis by \citet{Yang}, motivated by the presence of several unidentified 4FGL point sources in the region, revealed instead an extended diffuse emission component modeled as a disk with an extension of $\sim 0.6^\circ$, also mildly offset from the cluster. A map of the region is shown in Fig.~\ref{fig:w43_gamma_and_sources}. 

At radio wavelengths, \citet{Luque_Escamilla_w43_radio} analyzed the W43 region using VLA and NRAO data and identified compact non-thermal synchrotron emission likely associated with the cluster. The central radio condensation has a deconvolved angular size of approximately $4''\times4''$, while the full radio structure extends over roughly $15''\times15''$. 
Given its highly compact morphology, the radio emission may originate from shocks associated with colliding-wind binaries rather than from individual or collective wind termination shocks. Alternatively, it could also trace a very localized region of enhanced magnetic field or particle acceleration within a more extended wind interaction region.
The radio emission therefore indicates the presence of relativistic electrons, but its compact morphology and uncertain origin prevent a direct multi-wavelength flux comparison with the much more extended gamma-ray emission.

In the X-ray band, several studies have focused on the binarity of WR121a \citep{Anderson2011,Arora2020}, reporting an X-ray luminosity of $\sim 10^{34}\,\rm erg/s$. However, no dedicated extended X-ray analysis of the larger region has been published. Therefore, we used Chandra data to obtain an upper limit on the energy flux at a few keV over a $\sim 0.5^\circ$ region centered on the H.E.S.S. emission  (details are presented in Appendix~\ref{Appendix:X-rays}).

\subsection{Other possible counterpart and cloud/clump catalogs}

\begin{figure}
    \centering
    \includegraphics[width=0.9\columnwidth]{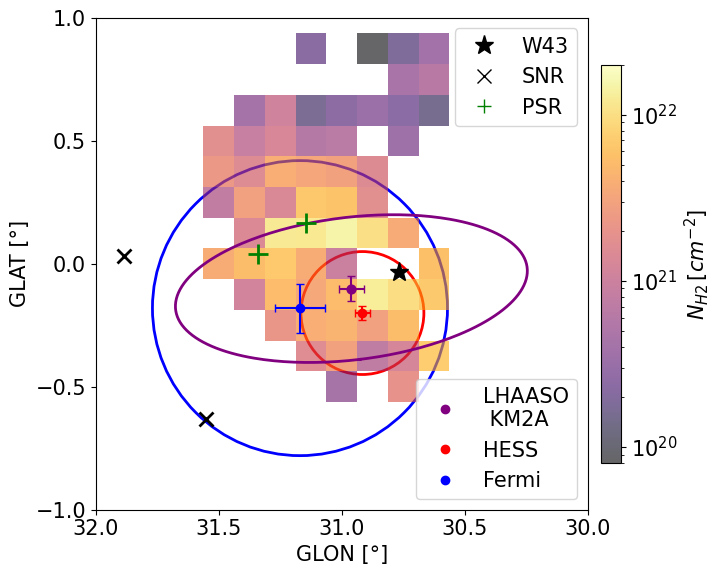}
    \caption{Map of the W43 region showing the centroids and extensions of
    the gamma-ray emission detected by Fermi-LAT \citep{Yang}, HESS \citep{HESS} and LHAASO KM2A \citep{Wang}. The gas column density of cloud $274$ from \citet{Miville-Deschenes} is also shown, together with the positions of the closest known bright pulsars and SNRs.}
    \label{fig:w43_gamma_and_sources}
\end{figure}

For completeness, we discuss other potential counterparts to the observed non-thermal emission. The closest isolated SNRs are G031.5-00.6, G031.9+00.0 \citep{green2025}, but they lie outside the gamma-ray emission region, with angular separations exceeding $\sim 1^\circ$, as indicated by gray crosses in Fig.~\ref{fig:w43_gamma_and_sources}.
Although these SNRs could in principle illuminate a MC coincident with the gamma-ray emission, their respective relatively large distances from the Sun of $\sim 13~\rm kpc$ \citep{Case1998} and $\sim 7~\rm kpc$ \citep{Ferrand2012}, together with the fact that a CR propagation distance of at least $\sim 100~\rm pc$ would be required based on the angular separation, disfavor this scenario.
Only two pulsars are also located inside the gamma-ray emission region, but their spin-down luminosity divided by distance squared seems too low ($\ll 10^{35}\,\rm erg\,s^{-1}\,kpc^{-2}$) to power a pulsar wind nebula capable of explaining the extended emission \citep{Acero2013}. Even scenarios involving past activity, where a pulsar has moved away from its birth site due to a kick velocity, appear disfavored given the required energetics and offsets. The only pulsar with a sufficiently high $\dot{E}/d^2 $ is PSR J1846-0258 (spatially coincident with SNR G029.7-00.3) but it is located $\sim 1.5^\circ$ away from the center of the gamma-ray emission, making an association unlikely.
Finally, it is possible that an undetected pulsar located near the emission center powers a pulsar wind nebula, with its radio or gamma-ray beam not intersecting the line of sight, as expected for a significant fraction of Galactic pulsars \citep{Shimasue2026}.
We also note that the globular cluster GLIMPSE 01 lies within the region and has an associated Fermi source. In the absence of clear evidence for these counterparts, we will assume in the following that the emission is effectively powered by the central W43-main stellar cluster.

Concerning the target gas within the gamma-ray emission region, \citet{Yang} estimated a total gas mass of $3.5 \times 10^6~M_{\odot}$, consistent with the values reported by \citet{Motte2} and \citet{Nguyen_2011_w43}. Additional information on the gas distribution, distances, and masses of individual molecular structures can be obtained from the catalog of \citet{Miville-Deschenes,MivilleDeschenes_data2021}. In particular, cloud $274$ exhibits a remarkable spatial correspondence with the gamma-ray morphology, as shown in Fig.~\ref{fig:w43_gamma_and_sources}, where column densities were derived using the conversion factor of \citet{Bolatto2013}.
Cloud $274$ has a mass of $\sim 1.2\times10^6~M_{\odot}$ and a near kinematic distance estimate of $6.7~\rm kpc$. 
Although this value is somewhat larger than the cluster distance of $5.5~\rm kpc$, kinematic distance estimates are typically affected by uncertainties at the level of several tens of percent. 
We therefore consider this cloud compatible with being illuminated by the W43 stellar cluster, and retain its observed value of $M/d^2\sim3 \times 10^4 \rm ~M_{\odot}/kpc^2$. 

Two other major molecular structures are present in the region, namely clouds $121$ and $546$, with masses of $1.6\times10^6~M_{\odot}$ and $8\times10^5~M_{\odot}$, and estimated distances of $5.75~\rm kpc$ and $4.9~\rm kpc$, respectively. Their spatial correspondence with the gamma-ray emission is less compelling than for cloud $274$, although they may still contribute, particularly to the more extended GeV emission observed by Fermi. Together, these three structures account for most of the molecular material observed toward the region and recover the total gas mass estimated by \citet{Yang}.
In the following, we adopt the value of $M/d^2$ derived for cloud $274$, as it provides the simplest interpretation involving a single molecular structure, but it should be kept in mind that the effective target mass could be larger by up to a factor of $\sim 3$ if multiple clouds contribute to the observed gamma-ray emission.

\subsection{Hadronic Modeling}

\begin{figure}
   \centering \includegraphics[width=0.9\columnwidth]{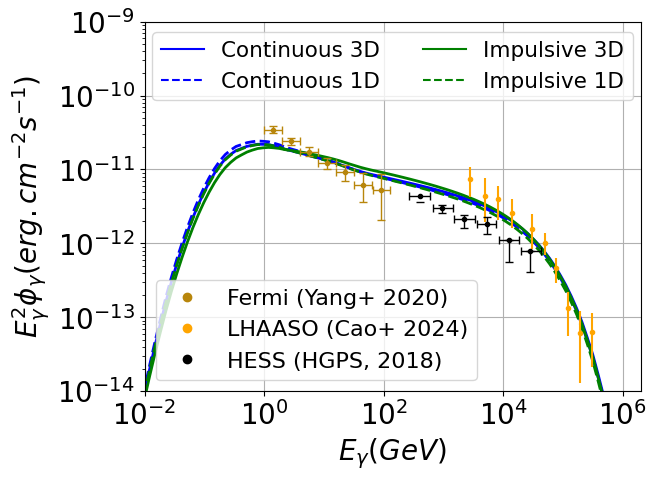}
   \caption{Model spectra matching the gamma-ray emission from Fermi \citep{Yang}, LHAASO \citep{Wang}, and H.E.S.S. \citep{HESS}.}
    \label{fig:w43_hadronic}%
 \end{figure}

Having argued that alternative counterparts are unlikely to account for the observed non-thermal emission, we now investigate whether the gamma-ray data can be explained by hadronic CRs escaping from the W43-main stellar cluster. To this end, we use the GeV spectrum derived from the Fermi analysis of \citet{Yang} together with the TeV measurements reported by \citet{Wang}. 
The H.E.S.S. data points lay below the LHAASO ones, probably because the angular extension of the H.E.S.S. source is significantly smaller (Fig.~\ref{fig:w43_hadronic}).
We consider both continuous injection, representing the combined contribution of stellar winds and successive SNRs as explained in Sec.~\ref{sec:mech_lum}, and impulsive injection, corresponding to the possibility of a recent SNR explosion within the cluster. This will allow us to determine which scenario provides the best description of the data and to constrain the associated physical parameters.
Moreover, we adopt the fiducial values derived in the previous sections, namely a total mechanical power $L_m\sim10^{38}$ erg/s, a cluster age $t\sim5~\rm Myr$, and a target gas parameter $M/d^2 \sim 3\times10^4~\rm M_{\odot}/kpc^2$. For the impulsive scenario, we assume a supernova explosion energy of $10^{51}~\rm erg$, while the CR acceleration efficiency $\eta$ is left as a free parameter in both cases. 

The distance between the accelerator and the target MC is not directly constrained because of projection effects and therefore remains a free parameter of the model. 
In the following, we adopt the plausible assumption that the MC is located near the outer regions of the wind bubble (partially surviving the interaction with the FS due to its high density; \citealt{Rogers2013}), and we set its distance to a fiducial value of $R=60~\rm pc$. 
Although the true cluster-cloud separation remains uncertain and could plausibly differ by a factor of $\sim 2$ in either direction, significantly smaller distances would place the cloud deep inside the wind bubble where its survival becomes problematic, while significantly larger distances would make it difficult to reproduce the observed gamma-ray emission. 

The Fermi and LHAASO spectral points are shown in Fig.~\ref{fig:w43_hadronic}.
To explain data, an enhancement above the Galactic CR sea is required, by a factor of $\sim2$-$10$ in the GeV range and $\sim30$-$50$ in the TeV range.
To interpret this CR excess, useful insights can already be obtained from Fig.~\ref{fig:adim_par_wind}. 
Indeed, the large CR enhancement implied by data requires correspondingly large values of $\Lambda_{\rm CR}$. In addition, owing to the large cluster age, the continuous injection model is expected to lie close to the steady-state regime. Thus, for continuous isotropic diffusion, Eq.~\ref{eq:Lambda_CR} and Fig.~\ref{fig:adim_par_wind} suggest that the diffusion coefficient must be suppressed by roughly one to two orders of magnitude relative to the Galactic value. Alternatively, Eq.~\ref{eq:Lambda_CR_1D} shows that larger diffusion coefficients very close to the Galactic value are expected with the anisotropic diffusion model.
In contrast, for the impulsive case Eq.~\ref{eq:Lambda_CR_imp}, Fig.~\ref{fig:adim_par_snr} and Eq.~\ref{eq:Lambda_CR_imp_1D} indicate that only a moderate suppression is expected in both isotropic and anisotropic cases.

Our predicted gamma-ray emission is shown in Fig.~\ref{fig:w43_hadronic}.
Solid and dashed lines correspond to 3D and 1D diffusion, respectively, while blue and green curves denote continuous (stationary) and impulsive injection. 
For the continuous injection model, the curves in the figure have been obtained for $\alpha=2.0$, $\delta=0.3$. For the 3D isotropic diffusion model $(\eta/0.1)(10^{28}~\rm cm^2~s^{-1}/D_0)\simeq30$, while for the 1D model  $(\eta/0.1)(10^{28}~\rm cm^2~s^{-1}/D_0)\simeq0.5$ and $L=100~\rm pc$. 
For impulsive injection, curves have been computed for $(10^{28}~\rm cm^2~s^{-1}/D_0)(10^4~\rm yr/t_{SN})\simeq 0.7$, $\alpha=2.0$, $\delta=0.24$ and $\eta=0.4$ for the 3D model while $(10^{28}~\rm cm^2~s^{-1}/D_0)(10^4~\rm yr/t_{SN})\simeq 10$, $\alpha=2.0$, $\delta=0.22$ and $\eta \simeq 0.03$ for the 1D model, again assuming $L=100~\rm pc$.
Furthermore, all models require a cutoff at $E_{\max}\sim0.4~\rm PeV$, consistent with the maximum energies expected for both SNRs \citep{Schure2013} and WTSs \citep{harer2025,Inventar2026}.

Although all the curves shown in Fig.~\ref{fig:w43_hadronic} seem to provide similarly good match to the data, the inferred parameters tend to disfavor the impulsive scenario. 
Indeed, the derived small value of $\delta$ is in tension with that inferred from Galactic CR observations \citep{Strong2007}, while adopting a smaller value of $\alpha$ would imply a larger $\delta$ but it would be in tension with the expectations of diffusive shock acceleration theory. 
Moreover, the required acceleration efficiency (or equivalently, the injected mechanical energy) deviates from the standard value of $\sim10\%$, being a factor $4$ higher in the 3D case and a factor of $3$ lower in the 1D case. This suggests that either scenario would require a SNR with somewhat atypical acceleration properties.
For more standard parameter values, impulsive models would typically either overpredict the low-energy emission because of their steeper spectra, or fail to reproduce part of the GeV emission. A superposition of several SNRs with different recent ages could naturally alleviate these issues, although at the expense of introducing additional free parameters. 
We also note that the lowest-energy Fermi data point is not perfectly reproduced by the models. This discrepancy should be interpreted with caution as the point lies close to the peak of the CR-sea emission, where uncertainties in background subtraction may be large for such complex region. A slightly higher background estimate would lower the inferred flux at this energy and could bring the data point into better agreement with the model predictions.

Overall, while an impulsive injection scenario cannot be fully ruled out, it requires slightly peculiar parameter values, making the continuous injection scenario mildly more favorable.
If the continuous injection scenario is considered, we can combine our results for the 1D and 3D diffusion to constrain the ratio of the CR acceleration efficiency to the diffusion coefficient in the range:
\begin{equation}
0.5\lesssim \frac{\eta}{0.1} \frac{10^{28}~\rm cm^2/s}{D_0}\lesssim 30
\end{equation}
We stress that this result is obtained for the adopted fiducial values of $R$, $L_m$, and $M/d^2$, and may vary by a factor of a few if these parameters are slightly modified within plausible ranges. 

Still, this suggests similar results to those inferred for isolated SNRs environments, where a suppression of the isotropic diffusion coefficient is often invoked to reproduce gamma-ray emission \citep[e.g.][]{Gabici_W28}.
This is sometimes explained by invoking the suppression of the diffusion coefficient due to CR streaming instability \citep{dangelo2016}, but its effectiveness is still debated \citep{recchia2022}.
However, also for the case of isolated SNRs, anisotropic diffusion may provide a simpler and more natural alternative to streaming instability \citep{Nava}.

\subsection{Leptonic modeling}

We now explore the possibility that the observed emission is leptonic rather than hadronic. 
A leptonic interpretation could in principle also involve electrons escaping from the stellar cluster and producing gamma-rays through Bremsstrahlung in the dense MC. However, such a scenario is expected to be disfavored, because electrons undergo stronger radiative losses during propagation,
that tend to confine them quite close to their accelerator, and because Bremsstrahlung is a quite ineffective radiation process.
We consider instead a scenario in which electrons produce gamma-rays within the wind bubble, close to the cluster WTS, through IC scattering on the stellar UV radiation field and the CMB. Indeed, diffuse direct and dust-scattered starlight are subdominant, as shown by \citet{harer2023,harer2025}, while the infrared contribution from the cluster's dust-reprocessed emission is also expected to be limited. This is because the low density of the wind bubble causes most of the dust reprocessing to occur far from the WTS, such that the infrared radiation is strongly diluted, and will be neglected in the following.
Of course in this case the spatial coincidence between gamma-ray emission and the MC would be just accidental (i.e. due to projection effects).

To assess the viability of the IC scenario, we first estimate the maximum distance that high-energy electrons can propagate before cooling and compare it with the observed spatial extension of the gamma-ray sources. As gamma-rays extend up to $\sim 100~\rm TeV$, at this energy, Klein-Nishina effects on UV are important and the cooling is expected to be dominated by the CMB. Using the Thomson scaling, the corresponding maximal electron energy is $\sim 150~\rm TeV$ \citep{Ghisellini_2013}. Adopting a diffusion coefficient $D \sim 10^{28}(E/\rm GeV)^{0.3}$ and using a cooling dominated by the CMB of timescale $t_{\rm cool}\sim 10 \rm ~kyr ~(150~\rm TeV)/E $ \citep{Rybicki1979}, the corresponding propagation scale is
$R_e \sim \sqrt{D(E=150~\rm TeV)\,t_{\rm cool}}\sim 100~\rm pc$,
which is sufficient to account for the observed extension. We therefore proceed with a more detailed leptonic modeling.

\begin{figure} 
    \centering
    \includegraphics[width=0.9\columnwidth]{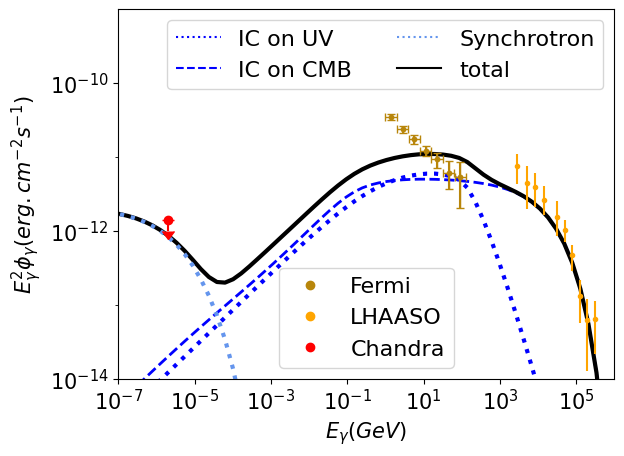}
    \caption{Inverse Compton scattering emission from electrons interacting with the CMB and UV ambient photons. Synchrotron emission from the same electrons is also shown. Details of the model are developed in the main text.}
    \label{fig:w43_gamma_IC}
\end{figure}

We assume a continuous injection of electrons in the cluster (from the WTS and/or successive SNRs), with a total available power $L_m \sim 10^{38}~\rm erg/s$, and a total injected energy $E_{\rm tot} = L_m \times t_{\rm age}$, of which $1\%$ is transferred to electrons. The ambient medium is taken to have a density of $\sim 0.1~\rm cm^{-3}$, representative of the bubble interior. Radiative losses include IC scattering on CMB and UV photon fields (including Klein-Nishina effects), as well as synchrotron emission.
For the UV field, we assume a characteristic temperature of $5\times 10^4~\rm K$, representative of massive star cluster radiation \citep{harer2023}. The UV luminosity is taken as the previously derived $L_{\rm bol}$, with an associated energy density estimated as $L/(4\pi R^2 c)$ at a characteristic distance of $10~\rm pc$ from the cluster, corresponding to the typical scale of the WTS. The resulting emission, computed with the GAMERA code \citep{Hahn2022}, is shown in Fig.~\ref{fig:w43_gamma_IC}.
The inferred injection spectrum is approximately $E^{-2}$, with a magnetic field of $B\sim 2~\mu\rm G$, to be consistent with the X-ray upper limits on synchrotron emission. 
A cutoff at $E_{\rm max}\sim 150~\rm TeV$ is required, consistent with the previous estimate and with the balance between the radiative cooling timescale and the acceleration timescale, as done in e.g \citet{harer2023}.

The IC scenario struggles to reproduce the lowest-energy Fermi data points, although this could potentially be alleviated by a two-zone model, with IC inside the bubble and Bremsstrahlung from escaped electrons. More generally, this scenario requires relatively optimistic assumptions, such as a high target photon energy density or an electron acceleration efficiency of $\sim 1\%$. 
The latter may be difficult to reconcile with lower electron-to-proton ratio ($K_{ep} \lesssim 10^{-3}$) inferred in other star clusters \citet{Peron2026}. 
In addition, the fact that the observed gamma-ray emission is not centered on the cluster represents another challenge for this scenario.

Finally, we note that secondary electrons produced in hadronic interactions within the cloud could also contribute, but this component is expected to be significantly weaker than the primary hadronic emission \citep{Gabici2009}, and is therefore unable to account for the observed flux.

Overall, the continuous hadronic scenario seems to provide the most consistent explanation of the data.

\section{Conclusions}
\label{sec:conclusion}

To summarize, we first derived analytical estimates for the CR excess as a function of distance from a source, depending on the main physical parameters (age, power, diffusion coefficient, etc.), for both continuous and impulsive injection scenarios. 
We also examined anisotropic diffusion and showed that, in both cases, it can effectively increase the possible CR excess compared to isotropic diffusion.

We then characterized the parameter space that leads to a given CR excess for generic sources, providing a framework directly applicable to different configurations and sources. In this context, we showed in particular that a quasi-steady state is typically reached for continuous injection under conditions relevant to stellar clusters.

For configurations producing an excess, we computed benchmark gamma-ray spectra gamma-ray of MCs irrradiated by CRs that escaped the cluster, and compared them with detector sensitivities. 
This was done for both isotropic and anisotropic diffusion, covering both time-dependent and steady-state regimes.

Finally, we applied this framework to the specific case of the W43 stellar cluster. In the case of isotropic diffusion, reproducing the data requires a suppressed diffusion coefficient, similar to what has been inferred in studies of isolated SNRs \citep[e.g.][]{Gabici_W28}. 
A similar suppression to explain this data has also been proposed by \citet{barreto2026}, who developed a mirror+scattering model in which magnetic mirroring and pitch-angle scattering reduce CR diffusion. Alternatively, anisotropic diffusion with a Galactic-like diffusion coefficient can also account for the observations and may provide a more natural explanation without requiring an effective reduction of the diffusion coefficient near sources. We also briefly examined leptonic scenarios for W43, finding them less favorable than hadronic ones.

We note that additional effects such as suppressed penetration of CRs into MCs could modify the results, particularly at low energies \citep{Yang2023}, while transport regimes outside the diffusion approximation (e.g. superdiffusion, \citealt{Lazarian2014,Lazarian2021}) could also be relevant and would require further development of this framework. We further emphasize that gamma-ray emission from illuminated clouds may account for some unidentified high-energy sources, as observed for example by LHAASO \citep{LHAASO_catalog}, and may also alleviate potential issues related to excessive grammage in scenarios where emission is attributed directly to the source region \citep{blasi2025}. Overall, this framework provides a flexible basis for future studies, including extensions to spatially varying diffusion coefficients and more detailed anisotropic transport models beyond simplified flux-tube descriptions.

      

\bibliographystyle{aa}
\bibliography{references.bib}

@ARTICLE{Uchiyama2013,
       author = {{Uchiyama}, Hideki and {Nobukawa}, Masayoshi and {Tsuru}, Takeshi Go and {Koyama}, Katsuji},
        title = "{K-Shell Line Distribution of Heavy Elements along the Galactic Plane Observed with Suzaku}",
      journal = {\pasj},
         year = 2013,
        month = feb,
       volume = {65},
          eid = {19},
        pages = {19},
          doi = {10.1093/pasj/65.1.19},
archivePrefix = {arXiv},
       eprint = {1209.0067},
 primaryClass = {astro-ph.HE},
       adsurl = {https://ui.adsabs.harvard.edu/abs/2013PASJ...65...19U}
}

@ARTICLE{Bartalucci2014,
       author = {{Bartalucci}, I. and {Mazzotta}, P. and {Bourdin}, H. and {Vikhlinin}, A.},
        title = "{Chandra ACIS-I particle background: an analytical model}",
      journal = {\aap},
         year = 2014,
        month = jun,
       volume = {566},
          eid = {A25},
        pages = {A25},
          doi = {10.1051/0004-6361/201423443},
archivePrefix = {arXiv},
       eprint = {1404.3587},
 primaryClass = {astro-ph.IM},
       adsurl = {https://ui.adsabs.harvard.edu/abs/2014A&A...566A..25B}
}

@ARTICLE{Wilm2000,
       author = {{Wilms}, J. and {Allen}, A. and {McCray}, R.},
        title = "{On the Absorption of X-Rays in the Interstellar Medium}",
      journal = {\apj},
         year = 2000,
        month = oct,
       volume = {542},
       number = {2},
        pages = {914-924},
          doi = {10.1086/317016},
archivePrefix = {arXiv},
       eprint = {astro-ph/0008425},
 primaryClass = {astro-ph},
       adsurl = {https://ui.adsabs.harvard.edu/abs/2000ApJ...542..914W}
}

@ARTICLE{Acero2013,
       author = {{Acero}, F. and {Ackermann}, M. and {Ajello}, M. and {Allafort}, A. and {Baldini}, L. and {Ballet}, J. and {Barbiellini}, G. and {Bastieri}, D. and {Bechtol}, K. and {Bellazzini}, R. and {Blandford}, R.~D. and {Bloom}, E.~D. and {Bonamente}, E. and {Bottacini}, E. and {Brandt}, T.~J. and {Bregeon}, J. and {Brigida}, M. and {Bruel}, P. and {Buehler}, R. and {Buson}, S. and {Caliandro}, G.~A. and {Cameron}, R.~A. and {Caraveo}, P.~A. and {Cecchi}, C. and {Charles}, E. and {Chaves}, R.~C.~G. and {Chekhtman}, A. and {Chiang}, J. and {Chiaro}, G. and {Ciprini}, S. and {Claus}, R. and {Cohen-Tanugi}, J. and {Conrad}, J. and {Cutini}, S. and {Dalton}, M. and {D'Ammando}, F. and {de Palma}, F. and {Dermer}, C.~D. and {Di Venere}, L. and {Silva}, E. do Couto e. and {Drell}, P.~S. and {Drlica-Wagner}, A. and {Falletti}, L. and {Favuzzi}, C. and {Fegan}, S.~J. and {Ferrara}, E.~C. and {Focke}, W.~B. and {Franckowiak}, A. and {Fukazawa}, Y. and {Funk}, S. and {Fusco}, P. and {Gargano}, F. and {Gasparrini}, D. and {Giglietto}, N. and {Giordano}, F. and {Giroletti}, M. and {Glanzman}, T. and {Godfrey}, G. and {Gr{\'e}goire}, T. and {Grenier}, I.~A. and {Grondin}, M.-H. and {Grove}, J.~E. and {Guiriec}, S. and {Hadasch}, D. and {Hanabata}, Y. and {Harding}, A.~K. and {Hayashida}, M. and {Hayashi}, K. and {Hays}, E. and {Hewitt}, J. and {Hill}, A.~B. and {Horan}, D. and {Hou}, X. and {Hughes}, R.~E. and {Inoue}, Y. and {Jackson}, M.~S. and {Jogler}, T. and {J{\'o}hannesson}, G. and {Johnson}, A.~S. and {Kamae}, T. and {Kawano}, T. and {Kerr}, M. and {Kn{\"o}dlseder}, J. and {Kuss}, M. and {Lande}, J. and {Larsson}, S. and {Latronico}, L. and {Lemoine-Goumard}, M. and {Longo}, F. and {Loparco}, F. and {Lovellette}, M.~N. and {Lubrano}, P. and {Marelli}, M. and {Massaro}, F. and {Mayer}, M. and {Mazziotta}, M.~N. and {McEnery}, J.~E. and {Mehault}, J. and {Michelson}, P.~F. and {Mitthumsiri}, W. and {Mizuno}, T. and {Monte}, C. and {Monzani}, M.~E. and {Morselli}, A. and {Moskalenko}, I.~V. and {Murgia}, S. and {Nakamori}, T. and {Nemmen}, R. and {Nuss}, E. and {Ohsugi}, T. and {Okumura}, A. and {Orienti}, M. and {Orlando}, E. and {Ormes}, J.~F. and {Paneque}, D. and {Panetta}, J.~H. and {Perkins}, J.~S. and {Pesce-Rollins}, M. and {Piron}, F. and {Pivato}, G. and {Porter}, T.~A. and {Rain{\`o}}, S. and {Rando}, R. and {Razzano}, M. and {Reimer}, A. and {Reimer}, O. and {Reposeur}, T. and {Ritz}, S. and {Roth}, M. and {Rousseau}, R. and {Saz Parkinson}, P.~M. and {Schulz}, A. and {Sgr{\`o}}, C. and {Siskind}, E.~J. and {Smith}, D.~A. and {Spandre}, G. and {Spinelli}, P. and {Suson}, D.~J. and {Takahashi}, H. and {Takeuchi}, Y. and {Thayer}, J.~G. and {Thayer}, J.~B. and {Thompson}, D.~J. and {Tibaldo}, L. and {Tibolla}, O. and {Tinivella}, M. and {Torres}, D.~F. and {Tosti}, G. and {Troja}, E. and {Uchiyama}, Y. and {Vandenbroucke}, J. and {Vasileiou}, V. and {Vianello}, G. and {Vitale}, V. and {Werner}, M. and {Winer}, B.~L. and {Wood}, K.~S. and {Yang}, Z.},
        title = "{Constraints on the Galactic Population of TeV Pulsar Wind Nebulae Using Fermi Large Area Telescope Observations}",
      journal = {\apj},
         year = 2013,
        month = aug,
       volume = {773},
       number = {1},
          eid = {77},
        pages = {77},
          doi = {10.1088/0004-637X/773/1/77},
archivePrefix = {arXiv},
       eprint = {1306.5735},
 primaryClass = {astro-ph.HE},
       adsurl = {https://ui.adsabs.harvard.edu/abs/2013ApJ...773...77A}
}

@ARTICLE{Addazi2022,
       author = {{Addazi}, A. and {Alvarez-Muniz}, J. and {Alves Batista}, R. and {Amelino-Camelia}, G. and {Antonelli}, V. and {Arzano}, M. and {Asorey}, M. and {Atteia}, J.-L. and {Bahamonde}, S. and {Bajardi}, F. and {Ballesteros}, A. and {Baret}, B. and {Barreiros}, D.~M. and {Basilakos}, S. and {Benisty}, D. and {Birnholtz}, O. and {Blanco-Pillado}, J.~J. and {Blas}, D. and {Bolmont}, J. and {Boncioli}, D. and {Bosso}, P. and {Calcagni}, G. and {Capozziello}, S. and {Carmona}, J.~M. and {Cerci}, S. and {Chernyakova}, M. and {Clesse}, S. and {Coelho}, J.~A.~B. and {Colak}, S.~M. and {Cortes}, J.~L. and {Das}, S. and {D'Esposito}, V. and {Demirci}, M. and {Di Luca}, M.~G. and {di Matteo}, A. and {Dimitrijevic}, D. and {Djordjevic}, G. and {Prester}, D. Dominis and {Eichhorn}, A. and {Ellis}, J. and {Escamilla-Rivera}, C. and {Fabiano}, G. and {Franchino-Vi{\~n}as}, S.~A. and {Frassino}, A.~M. and {Frattulillo}, D. and {Funk}, S. and {Fuster}, A. and {Gamboa}, J. and {Gent}, A. and {Gergely}, L. {\'A}. and {Giammarchi}, M. and {Giesel}, K. and {Glicenstein}, J.-F. and {Gracia-Bond{\'\i}a}, J. and {Gracia-Ruiz}, R. and {Gubitosi}, G. and {Guendelman}, E.~I. and {Gutierrez-Sagredo}, I. and {Haegel}, L. and {Heefer}, S. and {Held}, A. and {Herranz}, F.~J. and {Hinderer}, T. and {Illana}, J.~I. and {Ioannisian}, A. and {Jetzer}, P. and {Joaquim}, F.~R. and {Kampert}, K.-H. and {Uysal}, A. Karasu and {Katori}, T. and {Kazarian}, N. and {Kerszberg}, D. and {Kowalski-Glikman}, J. and {Kuroyanagi}, S. and {L{\"a}mmerzahl}, C. and {Said}, J. Levi and {Liberati}, S. and {Lim}, E. and {Lobo}, I.~P. and {L{\'o}pez-Moya}, M. and {Luciano}, G.~G. and {Manganaro}, M. and {Marcian{\`o}}, A. and {Mart{\'\i}n-Moruno}, P. and {Martinez}, Manel and {Martinez}, Mario and {Mart{\'\i}nez-Huerta}, H. and {Mart{\'\i}nez-Mirav{\'e}}, P. and {Masip}, M. and {Mattingly}, D. and {Mavromatos}, N. and {Mazumdar}, A. and {M{\'e}ndez}, F. and {Mercati}, F. and {Micanovic}, S. and {Mielczarek}, J. and {Miller}, A.~L. and {Milosevic}, M. and {Minic}, D. and {Miramonti}, L. and {Mitsou}, V.~A. and {Moniz}, P. and {Mukherjee}, S. and {Nardini}, G. and {Navas}, S. and {Niechciol}, M. and {Nielsen}, A.~B. and {Obers}, N.~A. and {Oikonomou}, F. and {Oriti}, D. and {Paganini}, C.~F. and {Palomares-Ruiz}, S. and {Pasechnik}, R. and {Pasic}, V. and {P{\'e}rez de los Heros}, C. and {Pfeifer}, C. and {Pieroni}, M. and {Piran}, T. and {Platania}, A. and {Rastgoo}, S. and {Relancio}, J.~J. and {Reyes}, M.~A. and {Ricciardone}, A. and {Risse}, M. and {Frias}, M.~D. Rodriguez and {Rosati}, G. and {Rubiera-Garcia}, D. and {Sahlmann}, H. and {Sakellariadou}, M. and {Salamida}, F. and {Saridakis}, E.~N. and {Satunin}, P. and {Schiffer}, M. and {Sch{\"u}ssler}, F. and {Sigl}, G. and {Sitarek}, J. and {Peracaula}, J. Sol{\`a} and {Sopuerta}, C.~F. and {Sotiriou}, T.~P. and {Spurio}, M. and {Staicova}, D. and {Stergioulas}, N. and {Stoica}, S. and {Stri{\v{s}}kovi{\'c}}, J. and {Stuttard}, T. and {Cerci}, D. Sunar and {Tavakoli}, Y. and {Ternes}, C.~A. and {Terzi{\'c}}, T. and {Thiemann}, T. and {Tinyakov}, P. and {Torri}, M.~D.~C. and {T{\'o}rtola}, M. and {Trimarelli}, C. and {Trze{\'s}niewski}, T. and {Tureanu}, A. and {Urban}, F.~R. and {Vagenas}, E.~C. and {Vernieri}, D. and {Vitagliano}, V. and {Wallet}, J.-C. and {Zornoza}, J.~D.},
        title = "{Quantum gravity phenomenology at the dawn of the multi-messenger era-A review}",
      journal = {Progress in Particle and Nuclear Physics},
         year = 2022,
        month = jul,
       volume = {125},
          eid = {103948},
        pages = {103948},
          doi = {10.1016/j.ppnp.2022.103948},
archivePrefix = {arXiv},
       eprint = {2111.05659},
 primaryClass = {hep-ph},
       adsurl = {https://ui.adsabs.harvard.edu/abs/2022PrPNP.12503948A}
}

@ARTICLE{aharon2007,
       author = {{Aharonian}, F. and {Akhperjanian}, A.~G. and {Bazer-Bachi}, A.~R. and {Beilicke}, M. and {Benbow}, W. and {Berge}, D. and {Bernl{\"o}hr}, K. and {Boisson}, C. and {Bolz}, O. and {Borrel}, V. and {Braun}, I. and {Brion}, E. and {Brown}, A.~M. and {B{\"u}hler}, R. and {B{\"u}sching}, I. and {Boutelier}, T. and {Carrigan}, S. and {Chadwick}, P.~M. and {Chounet}, L. -M. and {Coignet}, G. and {Cornils}, R. and {Costamante}, L. and {Degrange}, B. and {Dickinson}, H.~J. and {Djannati-Ata{\"\i}}, A. and {Drury}, L. O'c. and {Dubus}, G. and {Egberts}, K. and {Emmanoulopoulos}, D. and {Espigat}, P. and {Farnier}, C. and {Feinstein}, F. and {Ferrero}, E. and {Fiasson}, A. and {Fontaine}, G. and {Funk}, Seb. and {Funk}, S. and {F{\"u}{\ss}ling}, M. and {Gallant}, Y.~A. and {Giebels}, B. and {Glicenstein}, J.~F. and {Gl{\"u}ck}, B. and {Goret}, P. and {Hadjichristidis}, C. and {Hauser}, D. and {Hauser}, M. and {Heinzelmann}, G. and {Henri}, G. and {Hermann}, G. and {Hinton}, J.~A. and {Hoffmann}, A. and {Hofmann}, W. and {Holleran}, M. and {Hoppe}, S. and {Horns}, D. and {Jacholkowska}, A. and {de Jager}, O.~C. and {Kendziorra}, E. and {Kerschhaggl}, M. and {Kh{\'e}lifi}, B. and {Komin}, Nu. and {Kosack}, K. and {Lamanna}, G. and {Latham}, I.~J. and {Le Gallou}, R. and {Lemi{\`e}re}, A. and {Lemoine-Goumard}, M. and {Lohse}, T. and {Martin}, J.~M. and {Martineau-Huynh}, O. and {Marcowith}, A. and {Masterson}, C. and {Maurin}, G. and {McComb}, T.~J.~L. and {Moulin}, E. and {de Naurois}, M. and {Nedbal}, D. and {Nolan}, S.~J. and {Noutsos}, A. and {Olive}, J. -P. and {Orford}, K.~J. and {Osborne}, J.~L. and {Panter}, M. and {Pelletier}, G. and {Petrucci}, P. -O. and {Pita}, S. and {P{\"u}hlhofer}, G. and {Punch}, M. and {Ranchon}, S. and {Raubenheimer}, B.~C. and {Raue}, M. and {Rayner}, S.~M. and {Reimer}, A. and {Reimer}, O. and {Ripken}, J. and {Rob}, L. and {Rolland}, L. and {Rosier-Lees}, S. and {Rowell}, G. and {Sahakian}, V. and {Santangelo}, A. and {Saug{\'e}}, L. and {Schlenker}, S. and {Schlickeiser}, R. and {Schr{\"o}der}, R. and {Schwanke}, U. and {Schwarzburg}, S. and {Schwemmer}, S. and {Shalchi}, A. and {Sol}, H. and {Spangler}, D. and {Spanier}, F. and {Steenkamp}, R. and {Stegmann}, C. and {Superina}, G. and {Tam}, P.~H. and {Tavernet}, J. -P. and {Terrier}, R. and {Tluczykont}, M. and {van Eldik}, C. and {Vasileiadis}, G. and {Venter}, C. and {Vialle}, J.~P. and {Vincent}, P. and {V{\"o}lk}, H.~J. and {Wagner}, S.~J. and {Ward}, M.},
        title = "{Detection of extended very-high-energy {\ensuremath{\gamma}}-ray emission towards the young stellar cluster Westerlund 2}",
      journal = {\aap},
         year = 2007,
        month = jun,
       volume = {467},
       number = {3},
}

@ARTICLE{Aharonian2020,
       author = {{Aharonian}, Felix and {Peron}, Giada and {Yang}, Ruizhi and {Casanova}, Sabrina and {Zanin}, Roberta},
        title = "{Probing the sea of galactic cosmic rays with Fermi-LAT}",
      journal = {\prd},
         year = 2020,
        month = apr,
       volume = {101},
       number = {8},
          eid = {083018},
        pages = {083018},
          doi = {10.1103/PhysRevD.101.083018},
archivePrefix = {arXiv},
       eprint = {1811.12118},
 primaryClass = {astro-ph.HE},
       adsurl = {https://ui.adsabs.harvard.edu/abs/2020PhRvD.101h3018A}
}

@ARTICLE{Aharonian2024,
       author = {{Aharonian}, F. and {Benkhali}, F. Ait and {Aschersleben}, J. and {Ashkar}, H. and {Backes}, M. and {Martins}, V. Barbosa and {Batzofin}, R. and {Becherini}, Y. and {Berge}, D. and {Bernl{\"o}hr}, K. and {B{\"o}ttcher}, M. and {Bolmont}, J. and {de Bony de Lavergne}, M. and {Borowska}, J. and {Brose}, R. and {Brown}, A. and {Brun}, F. and {Bruno}, B. and {Burger-Scheidlin}, C. and {Casanova}, S. and {Celic}, J. and {Cerruti}, M. and {Chand}, T. and {Chandra}, S. and {Chen}, A. and {Chibueze}, J. and {Chibueze}, O. and {Cotter}, G. and {Cristofari}, P. and {Devin}, J. and {Djannati-Ata{\"\i}}, A. and {Djuvsland}, J. and {Dmytriiev}, A. and {Egberts}, K. and {Einecke}, S. and {Feijen}, K. and {Filipovic}, M. and {Fontaine}, G. and {Funk}, S. and {Gabici}, S. and {Gallant}, Y.~A. and {Glicenstein}, J.~F. and {Glombitza}, J. and {Grolleron}, G. and {Haerer}, L. and {He{\ss}}, B. and {Hinton}, J.~A. and {Hofmann}, W. and {Holch}, T.~L. and {Horns}, D. and {Huang}, Zhiqiu and {Jamrozy}, M. and {Jankowsky}, F. and {Jung-Richardt}, I. and {Kasai}, E. and {Katarzy{\'n}ski}, K. and {Khatoon}, R. and {Kh{\'e}lifi}, B. and {Klu{\'z}niak}, W. and {Komin}, Nu. and {Kosack}, K. and {Kostunin}, D. and {Kundu}, A. and {Lang}, R.~G. and {Le Stum}, S. and {Lemi{\`e}re}, A. and {Lemoine-Goumard}, M. and {Lenain}, J.-P. and {Leuschner}, F. and {Mackey}, J. and {Marandon}, V. and {Mart{\'\i}-Devesa}, G. and {Marx}, R. and {Mehta}, A. and {Mitchell}, A. and {Moderski}, R. and {Moghadam}, M.~O. and {Mohrmann}, L. and {Montanari}, A. and {Moulin}, E. and {de Naurois}, M. and {Niemiec}, J. and {Ohm}, S. and {Olivera-Nieto}, L. and {de Ona Wilhelmi}, E. and {Ostrowski}, M. and {Panny}, S. and {Pensec}, U. and {Peron}, G. and {P{\"u}hlhofer}, G. and {Quirrenbach}, A. and {Ravikularaman}, S. and {Regeard}, M. and {Reimer}, A. and {Reimer}, O. and {Ren}, H. and {Renaud}, M. and {Reville}, B. and {Rieger}, F. and {Rowell}, G. and {Rudak}, B. and {Ruiz-Velasco}, E. and {Sabri}, K. and {Sahakian}, V. and {Salzmann}, H. and {Santangelo}, A. and {Sasaki}, M. and {Sch{\"a}fer}, J. and {Sch{\"u}ssler}, F. and {Schutte}, H.~M. and {Sol}, H. and {Spencer}, S. and {Stawarz}, {\L}. and {Steinmassl}, S. and {Steppa}, C. and {Streil}, K. and {Sushch}, I. and {Taylor}, A.~M. and {Terrier}, R. and {Tsirou}, M. and {Tsuji}, N. and {van Eldik}, C. and {Vecchi}, M. and {Venter}, C. and {Vink}, J. and {Wagner}, S.~J. and {White}, R. and {Wierzcholska}, A. and {Zacharias}, M. and {Zdziarski}, A.~A. and {Zech}, A. and {{\.Z}ywucka}, N. and {H.~E.~S.~S. Collaboration}},
        title = "{Very-high-energy {\ensuremath{\gamma}}-Ray Emission from Young Massive Star Clusters in the Large Magellanic Cloud}",
      journal = {\apjl},
         year = 2024,
        month = jul,
       volume = {970},
       number = {1},
          eid = {L21},
        pages = {L21},
          doi = {10.3847/2041-8213/ad5e67},
archivePrefix = {arXiv},
       eprint = {2407.16219},
 primaryClass = {astro-ph.HE},
       adsurl = {https://ui.adsabs.harvard.edu/abs/2024ApJ...970L..21A}
}

@ARTICLE{Anderson2011,
       author = {{Anderson}, Gemma E. and {Gaensler}, B.~M. and {Kaplan}, David L. and {Posselt}, Bettina and {Slane}, Patrick O. and {Murray}, Stephen S. and {Mauerhan}, Jon C. and {Benjamin}, Robert A. and {Brogan}, Crystal L. and {Chakrabarty}, Deepto and {Drake}, Jeremy J. and {Drew}, Janet E. and {Grindlay}, Jonathan E. and {Hong}, Jaesub and {Lazio}, T. Joseph W. and {Lee}, Julia C. and {Steeghs}, Danny T.~H. and {van Kerkwijk}, Marten H.},
        title = "{Identification of a Population of X-ray-emitting Massive Stars in the Galactic Plane}",
      journal = {\apj},
         year = 2011,
        month = feb,
       volume = {727},
       number = {2},
          eid = {105},
        pages = {105},
          doi = {10.1088/0004-637X/727/2/105},
archivePrefix = {arXiv},
       eprint = {1011.3295},
 primaryClass = {astro-ph.HE},
       adsurl = {https://ui.adsabs.harvard.edu/abs/2011ApJ...727..105A}
}

@ARTICLE{Arora2020,
       author = {{Arora}, Bharti and {Pandey}, J.~C.},
        title = "{Unraveling the Nature of the Deeply Embedded Wolf-Rayet Star WR 121a}",
      journal = {\apj},
         year = 2020,
        month = mar,
       volume = {891},
       number = {2},
          eid = {104},
        pages = {104},
          doi = {10.3847/1538-4357/ab7337},
archivePrefix = {arXiv},
       eprint = {2002.02722},
 primaryClass = {astro-ph.HE},
       adsurl = {https://ui.adsabs.harvard.edu/abs/2020ApJ...891..104A}
}

@ARTICLE{Atoyan_Aharonian,
       author = {{Aharonian}, F.A. and {Atoyan}, A.M.},
        title = "{On the emissivity of {\ensuremath{\pi}}\^0\^-decay gamma radiation in the vicinity of accelerators of galactic cosmic rays.}",
      journal = {\aap},
         year = 1996,
        month = may,
       volume = {309},
        pages = {917-928},
       url = {https://ui.adsabs.harvard.edu/abs/1996A&A...309..917A},
}

@ARTICLE{Bally,
       author = {{Bally}, J. and {Anderson}, L.~D. and {Battersby}, C. and {Calzoletti}, L. and {Digiorgio}, A.~M. and {Faustini}, F. and {Ginsburg}, A. and {Li}, J.~Z. and {Nguyen Luong}, Q. and {Molinari}, S. and {Motte}, F. and {Pestalozzi}, M. and {Plume}, R. and {Rodon}, J. and {Schilke}, P. and {Schlingman}, W. and {Schneider-Bontemps}, N. and {Shirley}, Y. and {Stringfellow}, G.~S. and {Testi}, L. and {Traficante}, A. and {Veneziani}, M. and {Zavagno}, A.},
        title = "{Herschel observations of the W43 ``mini-starburst''}",
      journal = {\aap},
         year = 2010,
        month = jul,
       volume = {518},
          eid = {L90},
        pages = {L90},
          doi = {10.1051/0004-6361/201014596},
archivePrefix = {arXiv},
       eprint = {1005.4092},
 primaryClass = {astro-ph.GA},
       adsurl = {https://ui.adsabs.harvard.edu/abs/2010A&A...518L..90B}
}

@ARTICLE{Barreto2026,
       author = {{Barreto-Mota}, Lucas and {de Gouveia Dal Pino}, Elisabete M. and {Di Marco}, Gaetano and {Alves Batista}, Rafael and {Adduci Faria}, Stela},
        title = "{Cosmic Ray Diffusion and the Origin of Very High Energy Gamma-Ray Emission in Young Massive Stellar Clusters}",
      journal = {arXiv e-prints},
         year = 2026,
        month = aug,
          eid = {arXiv:2608.14547},
        pages = {arXiv:2608.14547},
          doi = {10.48550/arXiv.2608.14547},
archivePrefix = {arXiv},
       eprint = {2608.14547},
 primaryClass = {astro-ph.HE},
       adsurl = {https://ui.adsabs.harvard.edu/abs/2026arXiv260814547B}
}

@ARTICLE{blasi2025,
       author = {{Blasi}, Pasquale},
        title = "{Gamma rays from star clusters and implications for the origin of Galactic cosmic rays}",
      journal = {\aap},
         year = 2025,
        month = feb,
       volume = {694},
          eid = {A244},
        pages = {A244},
          doi = {10.1051/0004-6361/202453017},
archivePrefix = {arXiv},
       eprint = {2501.16097},
 primaryClass = {astro-ph.HE},
       adsurl = {https://ui.adsabs.harvard.edu/abs/2025A&A...694A.244B}
}

@ARTICLE{Blum,
       author = {{Blum}, R.~D. and {Damineli}, A. and {Conti}, P.~S.},
        title = "{The Stellar Content of Obscured Galactic Giant H II Regions. I. W43}",
      journal = {\aj},
         year = 1999,
        month = mar,
       volume = {117},
       number = {3},
        pages = {1392-1401},
          doi = {10.1086/300791},
archivePrefix = {arXiv},
       eprint = {astro-ph/9812070},
 primaryClass = {astro-ph},
       adsurl = {https://ui.adsabs.harvard.edu/abs/1999AJ....117.1392B}
}

@article{Bolatto2013,
  author  = {Bolatto, A. D. and Wolfire, M. and Leroy, A. K.},
  title   = {The CO-to-H$_2$ Conversion Factor},
  journal = {Annual Review of Astronomy and Astrophysics},
  year    = {2013},
  volume  = {51},
  number  = {1},
  pages   = {207--268},
  doi     = {10.1146/annurev-astro-082812-140944}
}

@ARTICLE{Bosch-Ramon2005,
       author = {{Bosch-Ramon}, V. and {Aharonian}, F.~A. and {Paredes}, J.~M.},
        title = "{Electromagnetic radiation initiated by hadronic jets from microquasars in the ISM}",
      journal = {\aap},
         year = 2005,
        month = mar,
       volume = {432},
       number = {2},
        pages = {609-618},
          doi = {10.1051/0004-6361:20041794},
archivePrefix = {arXiv},
       eprint = {astro-ph/0411508},
 primaryClass = {astro-ph},
       adsurl = {https://ui.adsabs.harvard.edu/abs/2005A&A...432..609B}
}

@article{Bykov2010,
   title={Cosmic ray current driven turbulence in shocks with efficient particle acceleration: the oblique, long-wavelength mode instability: CR current driven turbulence in shocks},
   volume={410},
   ISSN={0035-8711},
   url={http://dx.doi.org/10.1111/j.1365-2966.2010.17421.x},
   DOI={10.1111/j.1365-2966.2010.17421.x},
   number={1},
   journal={\mnras},
   publisher={Oxford University Press (OUP)},
   author={Bykov, A. M. and Osipov, S. M. and Ellison, D. C.},
   year={2010},
   month=sep, pages={39–52} }

@ARTICLE{Case1998,
       author = {{Case}, Gary L. and {Bhattacharya}, Dipen},
        title = "{A New {\ensuremath{\Sigma}}-D Relation and Its Application to the Galactic Supernova Remnant Distribution}",
      journal = {\apj},
         year = 1998,
        month = sep,
       volume = {504},
       number = {2},
        pages = {761-772},
          doi = {10.1086/306089},
archivePrefix = {arXiv},
       eprint = {astro-ph/9807162},
 primaryClass = {astro-ph},
       adsurl = {https://ui.adsabs.harvard.edu/abs/1998ApJ...504..761C}
}

@ARTICLE{Celli2023,
       author = {{Celli}, S. and {Specovius}, A. and {Menchiari}, S. and {Mitchell}, A. and {Morlino}, G.},
        title = "{Mass and wind luminosity of young Galactic open clusters in Gaia DR2}",
      journal = {\aap},
         year = 2024,
        month = jun,
       volume = {686},
          eid = {A118},
        pages = {A118},
          doi = {10.1051/0004-6361/202348541},
archivePrefix = {arXiv},
       eprint = {2311.09089},
 primaryClass = {astro-ph.GA},
       adsurl = {https://ui.adsabs.harvard.edu/abs/2024A&A...686A.118C}
}

@ARTICLE{Celli2024,
       author = {{Celli}, S. and {Peron}, G.},
        title = "{Detection prospects of very and ultra high-energy gamma rays from extended sources with ASTRI, CTA, and LHAASO}",
      journal = {\aap},
         year = 2024,
        month = sep,
       volume = {689},
          eid = {A258},
        pages = {A258},
          doi = {10.1051/0004-6361/202449837},
archivePrefix = {arXiv},
       eprint = {2403.03731},
 primaryClass = {astro-ph.HE},
       adsurl = {https://ui.adsabs.harvard.edu/abs/2024A&A...689A.258C}
}

@ARTICLE{Celli_SNR,
       author = {{Celli}, S. and {Morlino}, G. and {Gabici}, S. and {Aharonian}, F.~A.},
        title = "{Exploring particle escape in supernova remnants through gamma rays}",
      journal = {\mnras},
         year = 2019,
        month = dec,
       volume = {490},
       number = {3},
        pages = {4317-4333},
          doi = {10.1093/mnras/stz2897},
archivePrefix = {arXiv},
       eprint = {1906.09454},
 primaryClass = {astro-ph.HE},
       adsurl = {https://ui.adsabs.harvard.edu/abs/2019MNRAS.490.4317C}
}

@ARTICLE{cesmon1983,
       author = {{Cesarsky}, C.~J. and {Montmerle}, T.},
        title = "{Gamma-Rays from Active Regions in the Galaxy - the Possible Contribution of Stellar Winds}",
      journal = {\ssr},
         year = 1983,
        month = oct,
       volume = {36},
       number = {2},
        pages = {173-193},
          doi = {10.1007/BF00167503},
       adsurl = {https://ui.adsabs.harvard.edu/abs/1983SSRv...36..173C}
}

@INPROCEEDINGS{Chaves2008,
       author = {{Chaves}, R.~C.~G. and {Renaud}, M. and {Lemoine-Goumard}, M. and {Goret}, P.},
        title = "{HESS J1848-018: Discovery Of VHE {\ensuremath{\gamma}}-ray Emission From The Direction Of W 43}",
    booktitle = {American Institute of Physics Conference Series},
         year = 2008,
       editor = {{Aharonian}, Felix A. and {Hofmann}, Werner and {Rieger}, Frank},
       volume = {1085},
        month = dec,
    publisher = {AIP},
        pages = {372-375},
          doi = {10.1063/1.3076684},
       adsurl = {https://ui.adsabs.harvard.edu/abs/2008AIPC.1085..372C}
}

@ARTICLE{crowth2007,
       author = {{Crowther}, Paul A.},
        title = "{Physical Properties of Wolf-Rayet Stars}",
      journal = {\araa},
         year = 2007,
        month = sep,
       volume = {45},
       number = {1},
        pages = {177-219},
          doi = {10.1146/annurev.astro.45.051806.110615},
archivePrefix = {arXiv},
       eprint = {astro-ph/0610356},
 primaryClass = {astro-ph},
       adsurl = {https://ui.adsabs.harvard.edu/abs/2007ARA&A..45..177C}
}

@ARTICLE{dangelo2016,
       author = {{D'Angelo}, Marta and {Blasi}, Pasquale and {Amato}, Elena},
        title = "{Grammage of cosmic rays around Galactic supernova remnants}",
      journal = {\prd},
         year = 2016,
        month = oct,
       volume = {94},
       number = {8},
          eid = {083003},
        pages = {083003},
          doi = {10.1103/PhysRevD.94.083003},
archivePrefix = {arXiv},
       eprint = {1512.05000},
 primaryClass = {astro-ph.HE},
       adsurl = {https://ui.adsabs.harvard.edu/abs/2016PhRvD..94h3003D}
}

@ARTICLE{Ferrand2012,
       author = {{Ferrand}, Gilles and {Safi-Harb}, Samar},
        title = "{A census of high-energy observations of Galactic supernova remnants}",
      journal = {Advances in Space Research},
         year = 2012,
        month = may,
       volume = {49},
       number = {9},
        pages = {1313-1319},
          doi = {10.1016/j.asr.2012.02.004},
archivePrefix = {arXiv},
       eprint = {1202.0245},
 primaryClass = {astro-ph.HE},
       adsurl = {https://ui.adsabs.harvard.edu/abs/2012AdSpR..49.1313F}
}

@ARTICLE{Fujita_SNR,
       author = {{Fujita}, Yutaka and {Ohira}, Yutaka and {Tanaka}, Shuta J. and {Takahara}, Fumio},
        title = "{Molecular Clouds as a Probe of Cosmic-Ray Acceleration in a Supernova Remnant}",
      journal = {\apjl},
         year = 2009,
        month = dec,
       volume = {707},
       number = {2},
        pages = {L179-L183},
          doi = {10.1088/0004-637X/707/2/L179},
archivePrefix = {arXiv},
       eprint = {0911.4482},
 primaryClass = {astro-ph.HE},
       adsurl = {https://ui.adsabs.harvard.edu/abs/2009ApJ...707L.179F}
}

@ARTICLE{Gabici2009,
       author = {{Gabici}, S. and {Aharonian}, F.~A. and {Casanova}, S.},
        title = "{Broad-band non-thermal emission from molecular clouds illuminated by cosmic rays from nearby supernova remnants}",
      journal = {\mnras},
         year = 2009,
        month = jul,
       volume = {396},
       number = {3},
        pages = {1629-1639},
          doi = {10.1111/j.1365-2966.2009.14832.x},
archivePrefix = {arXiv},
       eprint = {0901.4549},
 primaryClass = {astro-ph.HE},
       adsurl = {https://ui.adsabs.harvard.edu/abs/2009MNRAS.396.1629G}
}

@INPROCEEDINGS{Gabici_W28,
       author = {{Gabici}, S. and {Casanova}, S. and {Aharonian}, F.~A. and {Rowell}, G.},
        title = "{Constraints on the cosmic ray diffusion coefficient in the W28 region from gamma-ray observations}",
    booktitle = {Proceedings of the Annual meeting of the French Society of Astronomy and Astrophysics},
         year = 2010,
       editor = {{Boissier}, S. and {Heydari-Malayeri}, M. and {Samadi}, R. and {Valls-Gabaud}, D.},
        month = dec,
        pages = {313},
          doi = {10.48550/arXiv.1009.5291},
archivePrefix = {arXiv},
       eprint = {1009.5291},
 primaryClass = {astro-ph.HE},
       adsurl = {https://ui.adsabs.harvard.edu/abs/2010sf2a.conf..313G}
}

@ARTICLE{Gabici_SNR,
       author = {{Gabici}, Stefano},
        title = "{Cosmic ray escape from supernova remnants}",
      journal = {\memsai},
         year = 2011,
        month = jan,
       volume = {82},
        pages = {760},
          doi = {10.48550/arXiv.1108.4844},
archivePrefix = {arXiv},
       eprint = {1108.4844},
 primaryClass = {astro-ph.HE},
       url = {https://ui.adsabs.harvard.edu/abs/2011MmSAI..82..760G},
}

@ARTICLE{Gabici_Aharonian,
       author = {{Gabici}, Stefano and {Aharonian}, Felix A.},
        title = "{Searching for Galactic Cosmic-Ray Pevatrons with Multi-TeV Gamma Rays and Neutrinos}",
      journal = {\apjl},
         year = 2007,
        month = aug,
       volume = {665},
       number = {2},
        pages = {L131-L134},
          doi = {10.1086/521047},
archivePrefix = {arXiv},
       eprint = {0705.3011},
 primaryClass = {astro-ph},
       url = {https://ui.adsabs.harvard.edu/abs/2007ApJ...665L.131G},
}

@INPROCEEDINGS{gabici2024,
       author = {{Gabici}, S.},
        title = "{Star clusters as cosmic ray accelerators}",
    booktitle = {7th Heidelberg International Symposium on High-Energy Gamma-Ray Astronomy},
         year = 2024,
        month = dec,
        pages = {16},
       adsurl = {https://ui.adsabs.harvard.edu/abs/2024hegr.confE..16G}
}

@ARTICLE{gabici2019,
       author = {{Gabici}, Stefano and {Evoli}, Carmelo and {Gaggero}, Daniele and {Lipari}, Paolo and {Mertsch}, Philipp and {Orlando}, Elena and {Strong}, Andrew and {Vittino}, Andrea},
        title = "{The origin of Galactic cosmic rays: Challenges to the standard paradigm}",
      journal = {International Journal of Modern Physics D},
         year = 2019,
        month = jan,
       volume = {28},
       number = {15},
          eid = {1930022-339},
        pages = {1930022-339},
          doi = {10.1142/S0218271819300222},
archivePrefix = {arXiv},
       eprint = {1903.11584},
 primaryClass = {astro-ph.HE},
       adsurl = {https://ui.adsabs.harvard.edu/abs/2019IJMPD..2830022G}
}

@book{Ghisellini_2013,
   title={Radiative Processes in High Energy Astrophysics},
   ISBN={9783319006123},
   ISSN={1616-6361},
   url={http://dx.doi.org/10.1007/978-3-319-00612-3},
   DOI={10.1007/978-3-319-00612-3},
   journal={Lecture Notes in Physics},
   publisher={Springer International Publishing},
   author={Ghisellini, Gabriele},
   year={2013} }

@BOOK{Ginzburg,
       author = {{Ginzburg}, V.~L. and {Syrovatskii}, S.~I.},
        title = "{The origin of cosmic rays}",
         year = 1969,
       adsurl = {https://ui.adsabs.harvard.edu/abs/1969ocr..book.....G}
}

@ARTICLE{green2025,
       author = {{Green}, D.~A.},
        title = "{An updated catalogue of 310 Galactic supernova remnants and their statistical properties}",
      journal = {Journal of Astrophysics and Astronomy},
         year = 2025,
        month = jan,
       volume = {46},
       number = {1},
          eid = {14},
        pages = {14},
          doi = {10.1007/s12036-024-10038-4},
archivePrefix = {arXiv},
       eprint = {2411.03367},
 primaryClass = {astro-ph.GA},
       adsurl = {https://ui.adsabs.harvard.edu/abs/2025JApA...46...14G}
}

@ARTICLE{Gupta2018,
       author = {{Gupta}, Siddhartha and {Nath}, Biman B. and {Sharma}, Prateek and {Eichler}, David},
        title = "{Lack of thermal energy in superbubbles: hint of cosmic rays?}",
      journal = {\mnras},
         year = 2018,
        month = jan,
       volume = {473},
       number = {2},
        pages = {1537-1553},
          doi = {10.1093/mnras/stx2427},
archivePrefix = {arXiv},
       eprint = {1705.10448},
 primaryClass = {astro-ph.GA},
       adsurl = {https://ui.adsabs.harvard.edu/abs/2018MNRAS.473.1537G}
}

@software{Hahn2022,
       author = {{Hahn}, Joachim and {Romoli}, Carlo and {Breuhaus}, Mischa},
        title = "{GAMERA: Source modeling in gamma astronomy}",
 howpublished = {Astrophysics Source Code Library, record ascl:2203.007},
         year = 2022,
        month = mar,
          eid = {ascl:2203.007},
archivePrefix = {ascl},
       eprint = {2203.007},
       adsurl = {https://ui.adsabs.harvard.edu/abs/2022ascl.soft03007H}
}

@ARTICLE{harer2023,
       author = {{H{\"a}rer}, Lucia K. and {Reville}, Brian and {Hinton}, Jim and {Mohrmann}, Lars and {Vieu}, Thibault},
        title = "{Understanding the TeV {\ensuremath{\gamma}}-ray emission surrounding the young massive star cluster Westerlund 1}",
      journal = {\aap},
         year = 2023,
        month = mar,
       volume = {671},
}

@ARTICLE{harer2025,
       author = {{H{\"a}rer}, L. and {Vieu}, T. and {Schulze}, F. and {Larkin}, C.~J.~K. and {Reville}, B.},
        title = "{Deciphering the gamma-ray emission in the Cygnus region}",
      journal = {\aap},
         year = 2025,
        month = nov,
       volume = {703},
}

@ARTICLE{HESS,
       author = {{H.~E.~S.~S. Collaboration} and {Abdalla}, H. and {Abramowski}, A. and {Aharonian}, F. and {Ait Benkhali}, F. and {Ang{\"u}ner}, E.~O. and {Arakawa}, M. and {Arrieta}, M. and {Aubert}, P. and {Backes}, M. and et al.},
        title = "{The H.E.S.S. Galactic plane survey}",
      journal = {\aap},
         year = 2018,
        month = apr,
       volume = {612},
}

@ARTICLE{Inventar2026,
       author = {{Inventar}, A. and {Peron}, G. and {Recchia}, S. and {Gabici}, S.},
        title = "{Wolf─Rayet stars as tracers of gamma-ray emission: Isolated stars and stellar clusters or associations}",
      journal = {\aap},
         year = 2026,
        month = jun,
       volume = {710},
          eid = {A373},
        pages = {A373},
          doi = {10.1051/0004-6361/202557720},
archivePrefix = {arXiv},
       eprint = {2605.10294},
 primaryClass = {astro-ph.HE},
       adsurl = {https://ui.adsabs.harvard.edu/abs/2026A&A...710A.373I}
}

@ARTICLE{Kafexhiu,
       author = {{Kafexhiu}, Ervin and {Aharonian}, Felix and {Taylor}, Andrew M. and {Vila}, Gabriela S.},
        title = "{Parametrization of gamma-ray production cross sections for p p interactions in a broad proton energy range from the kinematic threshold to PeV energies}",
      journal = {\prd},
         year = 2014,
        month = dec,
       volume = {90},
       number = {12},
          eid = {123014},
        pages = {123014},
          doi = {10.1103/PhysRevD.90.123014},
archivePrefix = {arXiv},
       eprint = {1406.7369},
 primaryClass = {astro-ph.HE},
       url = {https://ui.adsabs.harvard.edu/abs/2014PhRvD..90l3014K},
}

@article{Lazarian2014,
  author  = {Lazarian, A. and Yan, H.},
  title   = {Superdiffusion of cosmic rays: implications for cosmic ray acceleration},
  journal = {Astrophys. J.},
  volume  = {784},
  pages   = {38},
  year    = {2014}
}

@article{Lazarian2021,
  author  = {Lazarian, A. and Xu, S.},
  title   = {Diffusion of cosmic rays in {MHD} turbulence with magnetic mirrors},
  journal = {Astrophys. J.},
  volume  = {923},
  pages   = {53},
  year    = {2021}
}

@ARTICLE{Lemoine_2025,
       author = {{Lemoine-Goumard}, Marianne and {H{\"a}rer}, Lucia and {Mohrmann}, Lars and {Bernet}, Romain and {Hinton}, Jim and {Peron}, Giada and {Reville}, Brian and {Tibaldo}, Luigi and {Vieu}, Thibault},
        title = "{A cosmic-ray loaded nascent outflow driven by a massive star cluster}",
      journal = {Nature Communications},
         year = 2025,
        month = dec,
       volume = {16},
       number = {1},
          eid = {10820},
        pages = {10820},
          doi = {10.1038/s41467-025-65592-4},
       adsurl = {https://ui.adsabs.harvard.edu/abs/2025NatCo..1610820L}
}

@ARTICLE{LHAASO_catalog,
       author = {{Cao}, Zhen and {Aharonian}, F. and {An}, Q. and {Axikegu} and {Bai}, Y.~X. and {Bao}, Y.~W. and {Bastieri}, D. and {Bi}, X.~J. and {Bi}, Y.~J. and {Cai}, J.~T. and et al.},
        title = "{The First LHAASO Catalog of Gamma-Ray Sources}",
      journal = {\apjs},
         year = 2024,
        month = mar,
       volume = {271},
       number = {1},
          eid = {25},
        pages = {25},
          doi = {10.3847/1538-4365/acfd29},
archivePrefix = {arXiv},
       eprint = {2305.17030},
 primaryClass = {astro-ph.HE},
       adsurl = {https://ui.adsabs.harvard.edu/abs/2024ApJS..271...25C}
}

@ARTICLE{Lester1985,
       author = {{Lester}, D.~F. and {Dinerstein}, H.~L. and {Werner}, M.~W. and {Harvey}, P.~M. and {Evans}, II, N.~J. and {Brown}, R.~L.},
        title = "{Star formation in the inner galaxy : a far-infrared and radio study of two HII regions.}",
      journal = {\apj},
         year = 1985,
        month = sep,
       volume = {296},
        pages = {565-575},
          doi = {10.1086/163475},
       adsurl = {https://ui.adsabs.harvard.edu/abs/1985ApJ...296..565L}
}

@ARTICLE{LHAASO_cygnus,
       author = {{Lhaaso Collaboration}},
        title = "{An ultrahigh-energy {\ensuremath{\gamma}} -ray bubble powered by a super PeVatron}",
      journal = {Science Bulletin},
         year = 2024,
        month = feb,
       volume = {69},
       number = {4},
        pages = {449-457},
          doi = {10.1016/j.scib.2023.12.040},
archivePrefix = {arXiv},
       eprint = {2310.10100},
 primaryClass = {astro-ph.HE},
       adsurl = {https://ui.adsabs.harvard.edu/abs/2024SciBu..69..449L}
}

@ARTICLE{Li_Chen,
       author = {{Li}, Hui and {Chen}, Yang},
        title = "{{\ensuremath{\gamma}}-rays from molecular clouds illuminated by accumulated diffusive protons from supernova remnant W28}",
      journal = {\mnras},
         year = 2010,
        month = nov,
       volume = {409},
       number = {1},
        pages = {L35-L38},
          doi = {10.1111/j.1745-3933.2010.00944.x},
archivePrefix = {arXiv},
       eprint = {1009.0894},
 primaryClass = {astro-ph.HE},
       url = {https://ui.adsabs.harvard.edu/abs/2010MNRAS.409L..35L},
}

@ARTICLE{Limongi_chieffi,
       author = {{Limongi}, Marco and {Chieffi}, Alessandro},
        title = "{The Nucleosynthesis of $^{26}$Al and $^{60}$Fe in Solar Metallicity Stars Extending in Mass from 11 to 120 M$_{solar}$: The Hydrostatic and Explosive Contributions}",
      journal = {\apj},
         year = 2006,
        month = aug,
       volume = {647},
       number = {1},
        pages = {483-500},
          doi = {10.1086/505164},
archivePrefix = {arXiv},
       eprint = {astro-ph/0604297},
 primaryClass = {astro-ph},
       adsurl = {https://ui.adsabs.harvard.edu/abs/2006ApJ...647..483L}
}

@ARTICLE{Louvet_w43_herschel,
       author = {{Louvet}, F. and {Motte}, F. and {Hennebelle}, P. and {Maury}, A. and {Bonnell}, I. and {Bontemps}, S. and {Gusdorf}, A. and {Hill}, T. and {Gueth}, F. and {Peretto}, N. and {Duarte-Cabral}, A. and {Stephan}, G. and {Schilke}, P. and {Csengeri}, T. and {Nguyen Luong}, Q. and {Lis}, D.~C.},
        title = "{The W43-MM1 mini-starburst ridge, a test for star formation efficiency models}",
      journal = {\aap},
         year = 2014,
        month = oct,
       volume = {570},
          eid = {A15},
        pages = {A15},
          doi = {10.1051/0004-6361/201423603},
archivePrefix = {arXiv},
       eprint = {1404.4843},
 primaryClass = {astro-ph.SR},
       adsurl = {https://ui.adsabs.harvard.edu/abs/2014A&A...570A..15L}
}

@ARTICLE{Luque_Escamilla_w43_radio,
       author = {{Luque-Escamilla}, P.~L. and {Mu{\~n}oz-Arjonilla}, A.~J. and {S{\'a}nchez-Sutil}, J.~R. and {Mart{\'\i}}, J. and {Combi}, J.~A. and {S{\'a}nchez-Ayaso}, E.},
        title = "{Infrared and radio study of the W43 cluster. Resolved binaries and non-thermal emission}",
      journal = {\aap},
         year = 2011,
        month = aug,
       volume = {532},
          eid = {A92},
        pages = {A92},
          doi = {10.1051/0004-6361/201117296},
archivePrefix = {arXiv},
       eprint = {1107.2783},
 primaryClass = {astro-ph.HE},
       adsurl = {https://ui.adsabs.harvard.edu/abs/2011A&A...532A..92L}
}

@ARTICLE{malkov2013,
       author = {{Malkov}, M.~A. and {Diamond}, P.~H. and {Sagdeev}, R.~Z. and {Aharonian}, F.~A. and {Moskalenko}, I.~V.},
        title = "{Analytic Solution for Self-regulated Collective Escape of Cosmic Rays from Their Acceleration Sites}",
      journal = {\apj},
         year = 2013,
        month = may,
       volume = {768},
       number = {1},
          eid = {73},
        pages = {73},
          doi = {10.1088/0004-637X/768/1/73},
archivePrefix = {arXiv},
       eprint = {1207.4728},
 primaryClass = {astro-ph.HE},
       adsurl = {https://ui.adsabs.harvard.edu/abs/2013ApJ...768...73M}
}

@ARTICLE{marcowith2025,
       author = {{Marcowith}, A.},
        title = "{Cosmic rays escape from their sources}",
      journal = {Frontiers in Astronomy and Space Sciences},
         year = 2025,
        month = jan,
       volume = {11},
          eid = {1411076},
        pages = {1411076},
          doi = {10.3389/fspas.2024.1411076},
       adsurl = {https://ui.adsabs.harvard.edu/abs/2025FrASS..1111076M}
}

@ARTICLE{Menchiari2024,
       author = {{Menchiari}, S. and {Morlino}, G. and {Amato}, E. and {Bucciantini}, N. and {Beltr{\'a}n}, M.~T.},
        title = "{Cygnus OB2 as a test case for particle acceleration in young massive star clusters}",
      journal = {\aap},
         year = 2024,
        month = jun,
       volume = {686},
          eid = {A242},
        pages = {A242},
          doi = {10.1051/0004-6361/202348817},
archivePrefix = {arXiv},
       eprint = {2402.07784},
 primaryClass = {astro-ph.HE},
       adsurl = {https://ui.adsabs.harvard.edu/abs/2024A&A...686A.242M}
}

@article{Merten2026,
    author = "Merten, Lukas and Aerdker, Sophie and Peretti, Enrico",
    title = "{Time-dependent cosmic-ray escape from wind bubbles: hard spectra formation}",
    eprint = "2606.12390",
    archivePrefix = "arXiv",
    primaryClass = "astro-ph.HE",
    month = "6",
    year = "2026"
}

@ARTICLE{Mestre2021,
       author = {{Mestre}, Enrique and {de O{\~n}a Wilhelmi}, Emma and {Torres}, Diego F. and {Holch}, Tim Lukas and {Schwanke}, Ullrich and {Aharonian}, Felix and {Parkinson}, Pablo Saz and {Yang}, Ruizhi and {Zanin}, Roberta},
        title = "{Probing the hadronic nature of the gamma-ray emission associated with Westerlund 2}",
      journal = {\mnras},
         year = 2021,
        month = aug,
       volume = {505},
       number = {2},
        pages = {2731-2740},
          doi = {10.1093/mnras/stab1455},
archivePrefix = {arXiv},
       eprint = {2105.09155},
 primaryClass = {astro-ph.HE},
       adsurl = {https://ui.adsabs.harvard.edu/abs/2021MNRAS.505.2731M}
}

@ARTICLE{Miville-Deschenes,
       author = {{Miville-Desch{\^e}nes}, Marc-Antoine and {Murray}, Norman and {Lee}, Eve J.},
        title = "{Physical Properties of Molecular Clouds for the Entire Milky Way Disk}",
      journal = {\apj},
         year = 2017,
        month = jan,
       volume = {834},
       number = {1},
          eid = {57},
        pages = {57},
          doi = {10.3847/1538-4357/834/1/57},
archivePrefix = {arXiv},
       eprint = {1610.05918},
 primaryClass = {astro-ph.GA},
       adsurl = {https://ui.adsabs.harvard.edu/abs/2017ApJ...834...57M}
}

@dataset{MivilleDeschenes_data2021,
       author = {{Miville-Desch{\^e}nes}, Marc-Antoine},
        title = "{Miville-Deschenes et al. 2017 - Molecular Cloud Catalog}",
 howpublished = {Harvard Dataverse dataset},
         year = 2021,
        month = jan,
          doi = {10.7910/DVN/QR9CFW},
       adsurl = {https://ui.adsabs.harvard.edu/abs/2021harv.data..439M}
}

@article{Morlino,
   title={Particle acceleration in winds of star clusters},
   volume={504},
   ISSN={1365-2966},
   url={http://dx.doi.org/10.1093/mnras/stab690},
   DOI={10.1093/mnras/stab690},
   number={4},
   journal={Monthly Notices of the Royal Astronomical Society},
   publisher={Oxford University Press (OUP)},
   author={Morlino, G and Blasi, P and Peretti, E and Cristofari, P},
   year={2021},
   month=mar, pages={6096–6105} }

@ARTICLE{Motte,
       author = {{Motte}, F. and {Schilke}, P. and {Lis}, D.~C.},
        title = "{From Massive Protostars to a Giant H II Region: Submillimeter Imaging of the Galactic Ministarburst W43}",
      journal = {\apj},
         year = 2003,
        month = jan,
       volume = {582},
       number = {1},
        pages = {277-291},
          doi = {10.1086/344538},
archivePrefix = {arXiv},
       eprint = {astro-ph/0208519},
 primaryClass = {astro-ph},
       adsurl = {https://ui.adsabs.harvard.edu/abs/2003ApJ...582..277M}
}

@ARTICLE{Motte2,
       author = {{Motte}, F. and {Nguy{\^e}n Luong}, Q. and {Schneider}, N. and {Heitsch}, F. and {Glover}, S. and {Carlhoff}, P. and {Hill}, T. and {Bontemps}, S. and {Schilke}, P. and {Louvet}, F. and {Hennemann}, M. and {Didelon}, P. and {Beuther}, H.},
        title = "{The formation of the W43 complex: constraining its atomic-to-molecular transition and searching for colliding clouds}",
      journal = {\aap},
         year = 2014,
        month = nov,
       volume = {571},
          eid = {A32},
        pages = {A32},
          doi = {10.1051/0004-6361/201323001},
archivePrefix = {arXiv},
       eprint = {1404.4404},
 primaryClass = {astro-ph.GA},
       adsurl = {https://ui.adsabs.harvard.edu/abs/2014A&A...571A..32M}
}

@ARTICLE{murase_2,
       author = {{Ohira}, Yutaka and {Murase}, Kohta and {Yamazaki}, Ryo},
        title = "{Gamma-rays from molecular clouds illuminated by cosmic rays escaping from interacting supernova remnants}",
      journal = {\mnras},
         year = 2011,
        month = jan,
       volume = {410},
       number = {3},
        pages = {1577-1582},
          doi = {10.1111/j.1365-2966.2010.17539.x},
archivePrefix = {arXiv},
       eprint = {1007.4869},
 primaryClass = {astro-ph.HE},
       url = {https://ui.adsabs.harvard.edu/abs/2011MNRAS.410.1577O}
}

@ARTICLE{Nava,
       author = {{Nava}, L. and {Gabici}, S.},
        title = "{Anisotropic cosmic ray diffusion and gamma-ray production close to supernova remnants, with an application to W28}",
      journal = {\mnras},
         year = 2013,
        month = feb,
       volume = {429},
       number = {2},
        pages = {1643-1651},
          doi = {10.1093/mnras/sts450},
archivePrefix = {arXiv},
       eprint = {1211.1668},
 primaryClass = {astro-ph.HE},
       adsurl = {https://ui.adsabs.harvard.edu/abs/2013MNRAS.429.1643N}
}

@ARTICLE{nava2016,
       author = {{Nava}, L. and {Gabici}, S. and {Marcowith}, A. and {Morlino}, G. and {Ptuskin}, V.~S.},
        title = "{Non-linear diffusion of cosmic rays escaping from supernova remnants - I. The effect of neutrals}",
      journal = {\mnras},
         year = 2016,
        month = oct,
       volume = {461},
       number = {4},
        pages = {3552-3562},
          doi = {10.1093/mnras/stw1592},
archivePrefix = {arXiv},
       eprint = {1606.06902},
 primaryClass = {astro-ph.HE},
       adsurl = {https://ui.adsabs.harvard.edu/abs/2016MNRAS.461.3552N}
}

@ARTICLE{Nguyen_2011_w43,
       author = {{Nguyen Luong}, Q. and {Motte}, F. and {Schuller}, F. and {Schneider}, N. and {Bontemps}, S. and {Schilke}, P. and {Menten}, K.~M. and {Heitsch}, F. and {Wyrowski}, F. and {Carlhoff}, P. and {Bronfman}, L. and {Henning}, T.},
        title = "{W43: the closest molecular complex of the Galactic bar?}",
      journal = {\aap},
         year = 2011,
        month = may,
       volume = {529},
          eid = {A41},
        pages = {A41},
          doi = {10.1051/0004-6361/201016271},
archivePrefix = {arXiv},
       eprint = {1102.3460},
 primaryClass = {astro-ph.SR},
       adsurl = {https://ui.adsabs.harvard.edu/abs/2011A&A...529A..41N}
}

@ARTICLE{peron2024,
       author = {{Peron}, Giada and {Casanova}, Sabrina and {Gabici}, Stefano and {Baghmanyan}, Vardan and {Aharonian}, Felix},
        title = "{The contribution of winds from star clusters to the Galactic cosmic-ray population}",
      journal = {Nature Astronomy},
         year = 2024,
        month = apr,
       volume = {8},
        pages = {530-537},
          doi = {10.1038/s41550-023-02168-6},
archivePrefix = {arXiv},
       eprint = {2407.07509},
 primaryClass = {astro-ph.HE},
       adsurl = {https://ui.adsabs.harvard.edu/abs/2024NatAs...8..530P}
}

@ARTICLE{peron2025,
       author = {{Peron}, G. and {Menchiari}, S. and {Morlino}, G. and {Amato}, E.},
        title = "{Hadronic acceleration in the young star cluster NGC 6611 inside the M16 region unveiled by Fermi-LAT: Constraints on the acceleration efficiency}",
      journal = {\aap},
         year = 2025,
        month = nov,
       volume = {703},
          eid = {L8},
        pages = {L8},
          doi = {10.1051/0004-6361/202556564},
archivePrefix = {arXiv},
       eprint = {2507.13009},
 primaryClass = {astro-ph.HE},
       adsurl = {https://ui.adsabs.harvard.edu/abs/2025A&A...703L...8P}
}

@INPROCEEDINGS{PeronICRC2025,
       author = {{Peron}, G.},
        title = "{Star clusters in the gamma-ray sky}",
    booktitle = {39th International Cosmic Ray Conference},
         year = 2025,
        month = dec,
          eid = {1390},
        pages = {1390},
          doi = {10.48550/arXiv.2511.11432},
archivePrefix = {arXiv},
       eprint = {2511.11432},
 primaryClass = {astro-ph.HE},
       adsurl = {https://ui.adsabs.harvard.edu/abs/2025icrc.confE1390P}
}

@ARTICLE{Peron2026,
       author = {{Peron}, G. and {Bracco}, A. and {Morlino}, G. and {Mantovanini}, S. and {Amato}, E. and {Galli}, D. and {Padovani}, M.},
        title = "{A broadband view of the thermal and nonthermal emission from the embedded massive star cluster RCW 38}",
      journal = {\aap},
         year = 2026,
        month = oct,
       volume = {714},
          eid = {A145},
        pages = {A145},
          doi = {10.1051/0004-6361/202661550},
archivePrefix = {arXiv},
       eprint = {2606.21288},
 primaryClass = {astro-ph.HE},
       adsurl = {https://ui.adsabs.harvard.edu/abs/2026A&A...714A.145P}
}

@ARTICLE{Ptuskin2008,
       author = {{Ptuskin}, V.~S. and {Zirakashvili}, V.~N. and {Plesser}, A.~A.},
        title = "{Non-linear diffusion of cosmic rays}",
      journal = {Advances in Space Research},
         year = 2008,
        month = aug,
       volume = {42},
       number = {3},
        pages = {486-490},
          doi = {10.1016/j.asr.2007.12.007},
       adsurl = {https://ui.adsabs.harvard.edu/abs/2008AdSpR..42..486P}
}

@ARTICLE{recchia2022,
       author = {{Recchia}, S. and {Galli}, D. and {Nava}, L. and {Padovani}, M. and {Gabici}, S. and {Marcowith}, A. and {Ptuskin}, V. and {Morlino}, G.},
        title = "{Grammage of cosmic rays in the proximity of supernova remnants embedded in a partially ionized medium}",
      journal = {\aap},
         year = 2022,
        month = apr,
       volume = {660},
          eid = {A57},
        pages = {A57},
          doi = {10.1051/0004-6361/202142558},
archivePrefix = {arXiv},
       eprint = {2106.04948},
 primaryClass = {astro-ph.HE},
       adsurl = {https://ui.adsabs.harvard.edu/abs/2022A&A...660A..57R}
}

@ARTICLE{Recchia2024,
       author = {{Recchia}, S. and {Gabici}, S.},
        title = "{Origin of the spectral features observed in the cosmic-ray spectrum}",
      journal = {\aap},
         year = 2024,
        month = dec,
       volume = {692},
          eid = {A20},
        pages = {A20},
          doi = {10.1051/0004-6361/202349005},
archivePrefix = {arXiv},
       eprint = {2312.11397},
 primaryClass = {astro-ph.HE},
       adsurl = {https://ui.adsabs.harvard.edu/abs/2024A&A...692A..20R}
}

@ARTICLE{Rocamora2025,
       author = {{Rocamora}, M. and {Reimer}, A. and {Mart{\'\i}-Devesa}, G. and {Kissmann}, R.},
        title = "{Exploring non-thermal emission from the star-forming region NGC 3603 through a realistic modelling of its environment}",
      journal = {\aap},
         year = 2025,
        month = jul,
       volume = {699},
          eid = {A136},
        pages = {A136},
          doi = {10.1051/0004-6361/202452883},
archivePrefix = {arXiv},
       eprint = {2411.05206},
 primaryClass = {astro-ph.HE},
       adsurl = {https://ui.adsabs.harvard.edu/abs/2025A&A...699A.136R}
}

@ARTICLE{Rogers2013,
       author = {{Rogers}, H. and {Pittard}, J.~M.},
        title = "{Feedback from winds and supernovae in massive stellar clusters - I. Hydrodynamics}",
      journal = {\mnras},
         year = 2013,
        month = may,
       volume = {431},
       number = {2},
        pages = {1337-1351},
          doi = {10.1093/mnras/stt255},
archivePrefix = {arXiv},
       eprint = {1302.2443},
 primaryClass = {astro-ph.SR},
       adsurl = {https://ui.adsabs.harvard.edu/abs/2013MNRAS.431.1337R}
}

@ARTICLE{roscro2015,
       author = {{Rosslowe}, C.~K. and {Crowther}, P.~A.},
        title = "{Spatial distribution of Galactic Wolf-Rayet stars and implications for the global population}",
      journal = {\mnras},
         year = 2015,
        month = mar,
       volume = {447},
       number = {3},
        pages = {2322-2347},
          doi = {10.1093/mnras/stu2525},
archivePrefix = {arXiv},
       eprint = {1412.0699},
 primaryClass = {astro-ph.SR},
       adsurl = {https://ui.adsabs.harvard.edu/abs/2015MNRAS.447.2322R}
}

@book{Rybicki1979,
  author    = {Rybicki, G.~B. and Lightman, A.~P.},
  title     = {Radiative Processes in Astrophysics},
  publisher = {Wiley-Interscience},
  year      = {1979}
}

@ARTICLE{Saha2020,
       author = {{Saha}, L. and {Dom{\'\i}nguez}, A. and {Tibaldo}, L. and {Marchesi}, S. and {Ajello}, M. and {Lemoine-Goumard}, M. and {L{\'o}pez}, M.},
        title = "{Morphological and Spectral Study of 4FGL J1115.1-6118 in the Region of the Young Massive Stellar Cluster NGC 3603}",
      journal = {\apj},
         year = 2020,
        month = jul,
       volume = {897},
       number = {2},
          eid = {131},
        pages = {131},
          doi = {10.3847/1538-4357/ab9ac2},
archivePrefix = {arXiv},
       eprint = {2006.00274},
 primaryClass = {astro-ph.HE},
       adsurl = {https://ui.adsabs.harvard.edu/abs/2020ApJ...897..131S}
}

@ARTICLE{Salpeter1955,
       author = {{Salpeter}, Edwin E.},
        title = "{The Luminosity Function and Stellar Evolution.}",
      journal = {\apj},
         year = 1955,
        month = jan,
       volume = {121},
        pages = {161},
          doi = {10.1086/145971},
       adsurl = {https://ui.adsabs.harvard.edu/abs/1955ApJ...121..161S}
}

@ARTICLE{Schure2013,
       author = {{Schure}, K.~M. and {Bell}, A.~R.},
        title = "{Cosmic ray acceleration in young supernova remnants}",
      journal = {\mnras},
         year = 2013,
        month = oct,
       volume = {435},
       number = {2},
        pages = {1174-1185},
          doi = {10.1093/mnras/stt1371},
archivePrefix = {arXiv},
       eprint = {1307.6575},
 primaryClass = {astro-ph.HE},
       adsurl = {https://ui.adsabs.harvard.edu/abs/2013MNRAS.435.1174S}
}

@article{Seo,
  author    = {Seo and KangHyesung and RyuDongsu},
  
  title     = {THE CONTRIBUTION OF STELLAR WINDS TO COSMIC RAY PRODUCTION},
  
  journal   = {Journal of The Korean Astronomical Society},
  
  volume    = {51},
  
  number    = {2},
  
  pages     = {37-48},
  year      = {2018},
}

@ARTICLE{Shimasue2026,
       author = {{Shimasue}, Takumi and {Kisaka}, Shota and {Bamba}, Aya and {Shibata}, Shinpei},
        title = "{Constraining the Pulsar Beaming Fraction with TeV-Selected Galactic Pulsar Wind Nebulae and unidentified TeV Sources}",
      journal = {arXiv e-prints},
         year = 2026,
        month = apr,
          eid = {arXiv:2604.16187},
        pages = {arXiv:2604.16187},
          doi = {10.48550/arXiv.2604.16187},
archivePrefix = {arXiv},
       eprint = {2604.16187},
 primaryClass = {astro-ph.HE},
       adsurl = {https://ui.adsabs.harvard.edu/abs/2026arXiv260416187S}
}

@ARTICLE{Smith,
       author = {{Smith}, L.~F. and {Biermann}, P. and {Mezger}, P.~G.},
        title = "{Star formation rates in the Galaxy.}",
      journal = {\aap},
         year = 1978,
        month = may,
       volume = {66},
        pages = {65-76},
       adsurl = {https://ui.adsabs.harvard.edu/abs/1978A&A....66...65S}
}

@article{Strong2007,
   title={Cosmic-Ray Propagation and Interactions in the Galaxy},
   volume={57},
   ISSN={1545-4134},
   url={http://dx.doi.org/10.1146/annurev.nucl.57.090506.123011},
   DOI={10.1146/annurev.nucl.57.090506.123011},
   number={1},
   journal={Annual Review of Nuclear and Particle Science},
   publisher={Annual Reviews},
   author={Strong, Andrew W. and Moskalenko, Igor V. and Ptuskin, Vladimir S.},
   year={2007},
   month=nov, pages={285–327} }

@ARTICLE{Torres_IC443,
       author = {{Torres}, Diego F. and {Marrero}, Ana Y. Rodr{\'\i}guez and {de Cea Del Pozo}, Elsa},
        title = "{The GeV to TeV connection in the environment of SNR IC 443}",
      journal = {\mnras},
         year = 2010,
        month = oct,
       volume = {408},
       number = {2},
}

@ARTICLE{Vacca1994,
       author = {{Vacca}, William D.},
        title = "{On the Determination of the Number of O Stars in H II Regions and Starburst Galaxies}",
      journal = {\apj},
         year = 1994,
        month = jan,
       volume = {421},
        pages = {140},
          doi = {10.1086/173632},
       adsurl = {https://ui.adsabs.harvard.edu/abs/1994ApJ...421..140V}
}

@ARTICLE{Vieu_superbubble,
       author = {{Vieu}, T. and {Gabici}, S. and {Tatischeff}, V. and {Ravikularaman}, S.},
        title = "{Cosmic ray production in superbubbles}",
      journal = {\mnras},
         year = 2022,
        month = may,
       volume = {512},
       number = {1},
}

@ARTICLE{Wang,
       author = {{LHAASO Collaboration}},
        title = "{Observation of the $\gamma$-ray Emission from W43 with LHAASO}",
      journal = {arXiv e-prints},
         year = 2024,
        month = aug,
          eid = {arXiv:2408.09905},
        pages = {arXiv:2408.09905},
          doi = {10.48550/arXiv.2408.09905},
archivePrefix = {arXiv},
       eprint = {2408.09905},
 primaryClass = {astro-ph.HE},
       adsurl = {https://ui.adsabs.harvard.edu/abs/2024arXiv240809905L}
}

@ARTICLE{weaver,
       author = {{Weaver}, R. and {McCray}, R. and {Castor}, J. and {Shapiro}, P. and {Moore}, R.},
        title = "{Interstellar bubbles. II. Structure and evolution.}",
      journal = {\apj},
         year = 1977,
        month = dec,
       volume = {218},
}

@ARTICLE{Yang2023,
       author = {{Yang}, Rui-zhi and {Li}, Guang-Xing and {Wilhelmi}, Emma de O{\~n}a and {Cui}, Yu-Dong and {Liu}, Bing and {Aharonian}, Felix},
        title = "{Effective shielding of {\ensuremath{\lesssim}}10 GeV cosmic rays from dense molecular clumps}",
      journal = {Nature Astronomy},
         year = 2023,
        month = mar,
       volume = {7},
        pages = {351-358},
          doi = {10.1038/s41550-022-01868-9},
archivePrefix = {arXiv},
       eprint = {2301.06716},
 primaryClass = {astro-ph.HE},
       adsurl = {https://ui.adsabs.harvard.edu/abs/2023NatAs...7..351Y}
}

@ARTICLE{Yang,
       author = {{Yang}, Rui-Zhi and {Wang}, Yuan},
        title = "{The diffuse gamma-ray emission toward the Galactic mini starburst W43}",
      journal = {\aap},
         year = 2020,
        month = aug,
       volume = {640},
          eid = {A60},
        pages = {A60},
          doi = {10.1051/0004-6361/202037518},
archivePrefix = {arXiv},
       eprint = {2007.15295},
 primaryClass = {astro-ph.HE},
       adsurl = {https://ui.adsabs.harvard.edu/abs/2020A&A...640A..60Y}
}

@article{Zhang,
   title={THE PARALLAX OF W43: A MASSIVE STAR-FORMING COMPLEX NEAR THE GALACTIC BAR},
   volume={781},
   ISSN={1538-4357},
   url={http://dx.doi.org/10.1088/0004-637X/781/2/89},
   DOI={10.1088/0004-637x/781/2/89},
   number={2},
   journal={\apj},
   publisher={American Astronomical Society},
   author={Zhang, B. and Moscadelli, L. and Sato, M. and Reid, M. J. and Menten, K. M. and Zheng, X. W. and Brunthaler, A. and Dame, T. M. and Xu, Y. and Immer, K.},
   year={2014},
   month=jan, pages={89} }

\begin{appendix}

\section{Time dependent solutions for 1D transport with free escape boundary}

\subsection{Impulsive injection}
\label{sec:app_eq_imp}

The solution of the 1D transport equation (Eq~\ref{eq:1D}) for impulsive injection was given in Eq~\ref{eq:anisotropic_SNR}.
However, as a magnetic flux tube has a finite size $L$, this solution is accurate only when $R_d<<L$. To extend it to arbitrary positions $R<L$, we can impose a free escape boundary ($n = 0$) at $R=L$.
Using the image method with one mirror located at $R=2L$, the following approximate solution is obtained:

\begin{equation}
\label{eq:1D_snr_time}
    n(t,E,z)\simeq\frac{N(E)}{2 \pi ^{3/2} R_{S}^2\sqrt{D(E)t} }\biggl[ \exp \left(-\frac{z^2}{4D(E)t}\right)-  \exp \left(-\frac{(2L-z)^2}{4D(E)t} \right)\biggr] 
\end{equation}

\subsection{Continuous injection}
\label{sec:app_eq_cont}

The solution of the 1D transport equation for continuous injection in a steady state regime with a free escape boundary at $R=L$ was given in Eq.\ref{eq:anisotropic_wind}. 
To obtain the full time-dependent solution, one must first solve Eq.~\ref{eq:1D} (changing the last term by $Q(E)/(\pi R_s^2) \times \delta (z)$) to obtain:

\begin{equation}
    n(t,E,z)=\frac{Q(E)}{2\pi D(E)R_{S}^2} \biggl[2 \biggl(\frac{Dt}{\pi}\biggr)^{1/2} e^{-\dfrac{z^2}{4Dt}} -z \cdot \mathrm{erfc}\biggl(\frac{z}{2\sqrt{Dt}} \biggr)\biggr]
\end{equation}

\noindent 
As for the impulsive case, the free escape boundary condition can now be enforced by applying the image method with one mirror located at $R=2L$, leading to the following approximate solution:

\begin{equation}
\label{eq:1D_wind_time}
    \begin{split}
        &  n(t,E,z)\simeq\frac{Q(E)}{4\pi D(E)R_{S}^2} \biggl[2 \biggl(\frac{Dt}{\pi}\biggr)^{1/2} e^{-\dfrac{z^2}{4Dt}} -z \cdot \mathrm{erfc}\biggl(\frac{z}{2\sqrt{Dt}} \biggr) \\
        &-2 \biggl(\frac{Dt}{\pi}\biggr)^{1/2} e^{-\dfrac{(2L-z)^2}{4Dt}} +(2L-z) \cdot \mathrm{erfc}\biggl(\frac{2L-z}{2\sqrt{Dt}} \biggr) \biggr] 
    \end{split}
\end{equation}

\section{Non-steady state characterization for continuous injection}
\label{app:nonsteady}

In Sec.~\ref{sec:plot_gamma_2} we studied the  time dependence of the gamma-ray emission from a MC illuminated by CRs escaping from a continuous source.
That was done for a fixed set of parameter values, and here we investigate the impact of varying such parameters.

\begin{figure} 
    \centering
    \includegraphics[width=0.95\columnwidth]{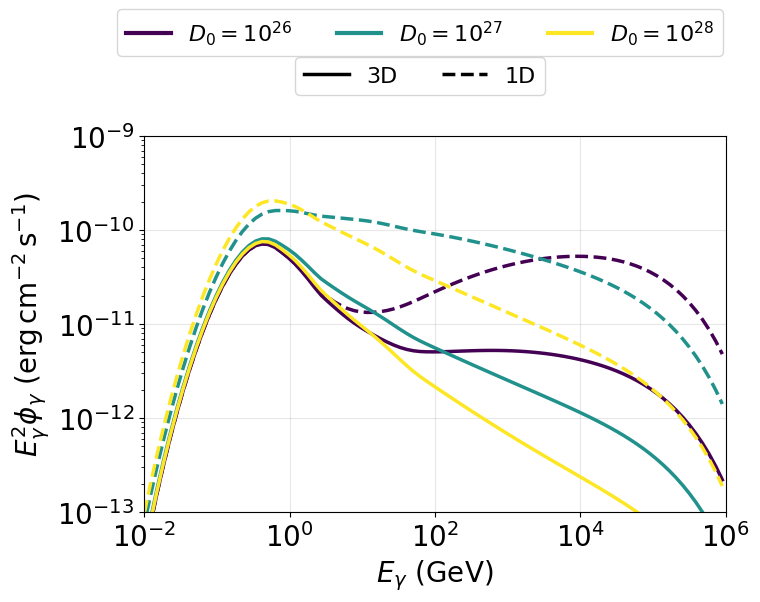}
    \caption{Same than Fig.~\ref{fig:gamma_wind_time} but at a fixed time $t=10^5 ~\rm yrs$ and with different colors representing different diffusion coefficient normalization (in units of $\rm cm^2/s$).}
    \label{fig:fluw_gamma_D0}
\end{figure}

\begin{figure} 
    \centering
    \includegraphics[width=0.95\columnwidth]{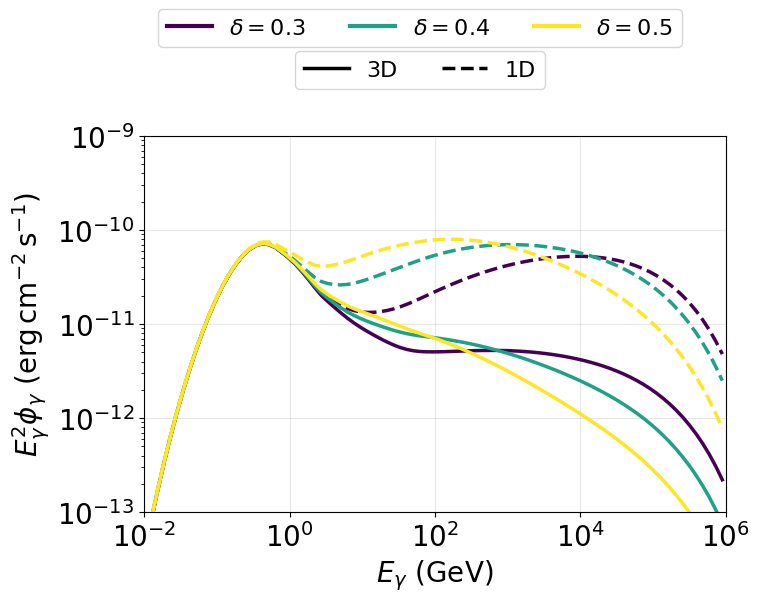}
    \caption{Same than Fig.~\ref{fig:gamma_wind_time} but at a fixed time $t=10^5 ~\rm yrs$ and with different colors representing different diffusion coefficient energy slopes.}
    \label{fig:fluw_gamma_delta}
\end{figure}

\begin{figure}
    \centering
    \includegraphics[width=0.95\columnwidth]{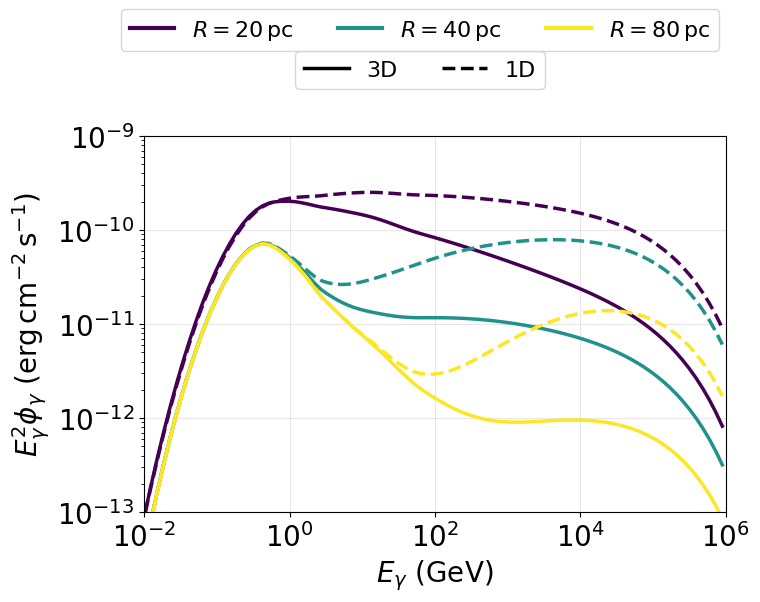}
    \caption{Same than Fig.~\ref{fig:gamma_wind_time} but at a fixed time $t=10^5 ~\rm yrs$ and with different colors representing different source-cloud distance.}
    \label{fig:fluw_gamma_radius}
\end{figure}

Fig.~\ref{fig:fluw_gamma_D0} shows the gamma-ray spectra obtained by varying the normalization of the diffusion coefficient $D_0$ (from $10^{26}$ to $10^{28}$ $\rm cm^2/s$), while keeping all other relevant parameters fixed, while Fig.~\ref{fig:fluw_gamma_delta} shows the gamma-ray spectra obtained by varying the energy dependence of the diffusion coefficient $\delta$ (from 0.3 to 0.5)
Finally, Fig.~\ref{fig:fluw_gamma_radius} shows the gamma-ray spectra obtained by varying the source-cloud distance $R$ (from 20 pc to 80 pc), while keeping all other relevant parameters fixed. 
In all cases, varying parameters can significantly impact on the results.

\vspace{2cm}

\section{Impulsive injection scenario}
\label{app:impulsive}

\subsection{Parameter space study}

Following Sec.~\ref{sec:CR_plots}, the CR excess at fixed energy in the impulsive scenario is also fully determined by the three quantities $\Lambda_{CR}^{imp}$, $\Gamma_{CR}^{imp}$, and $s^{imp}$ that allow us to rewrite Eq.~\ref{eq:impulsive} as
\begin{equation}
    n \approx 3\times 10^{-5} ~\Lambda_{CR}^{imp}~ (E/\rm GeV)^{-s^{imp}} ~\mathrm{exp}(-18/\Gamma^{imp}_{CR}) ~ \rm erg^{-1}cm^{-3}
\end{equation}
where
\begin{eqnarray}
\label{eq:Lambda_CR_imp}
    \Lambda_{CR}^{imp} &=&\frac{\alpha-2}{0.2}
~\frac{W_{CR}}{10^{50}\,\rm erg}
~\left (\frac{10^{28}\,\rm cm^2/s}{D_0}
~ \frac{10^3 ~\rm yrs}{t}\right)^{3/2} \\
\label{eq:Gamma_CR_imp}
    \Gamma_{CR}^{imp}&=&\frac{D_0}{10^{28}\,\rm cm^2/s}
~ \left ( \frac{E}{\rm GeV}\right)^{\delta}
~\frac{t}{10^3\,\rm yr}
~\left(\frac{50\,\rm pc}{R}\right)^2\\
s^{imp} &=& \alpha+\frac{3}{2}\delta 
\end{eqnarray}

The values of $\Lambda_{CR}^{imp}$ and $\Gamma_{CR}^{imp}$ that result in an excess of CRs above the Galactic sea of $n/n_{LIS} =$~1, 10, and 100 are plotted in Fig.~\ref{fig:adim_par_snr}. The three panels refer to different particle energies, and solid and dashed lines refer to two different values of $\alpha+\delta$.

The resulting scalings and overall behavior remain qualitatively similar than in the continuous case, even though the values of $\Gamma_{CR}^{imp}$ and $\Lambda_{CR}^{imp}$ are different. However, since both the normalization and the exponential argument depend explicitly on time, the x-axis can no longer be interpreted straightforwardly as a pure time-evolution parameter in the impulsive case. In addition, as it can be seen from the value of $\Lambda_{CR}^{imp}$, here a suppressed diffusion coefficient is not necessarily required to produce a significant excess, depending on time and distance.

Finally, from Eq.~\ref{eq:1D_snr_time}, a similar limitation than in the continuous scenario applies in the 1D case, where such a parameter-space representation is not possible due to the higher complexity of the solution. Nevertheless, a corresponding normalization factor and slope can still be defined for comparison, by re-writing the approximate solution of Eq.~\ref{eq:anisotropic_SNR} as 
\begin{equation}
    n \approx 4\times 10^{-5} ~\Lambda_{CR,~1D}^{imp}~ (E/\rm GeV)^{-s^{imp}_{1D}} ~\mathrm{exp}(-18/\Gamma^{imp}_{CR}) ~ \rm erg^{-1}cm^{-3} 
\end{equation}
where
\begin{align}
\label{eq:Lambda_CR_imp_1D}
\Lambda_{CR,~1D}^{imp} &= \frac{\alpha-2}{0.2}
~\frac{W_{CR}}{10^{50}\,\rm erg}
~\left (\frac{10^{28}\,\rm cm^2/s}{D_0}
~ \frac{10^3 ~\rm yrs}{t}\right)^{1/2}
\left(\frac{10\,\rm pc}{R_s}\right)^2 \\
s^{imp}_{1D} &= \alpha+\frac{1}{2}\delta 
\end{align}
essentially leading to a flux higher than in the isotropic case only at high energies, as the slope $\alpha + \tfrac{1}{2}\delta$ is harder than in the 3D case.

\subsection{Molecular clouds and gamma-rays}

Complementing Sec.~\ref{sec:plot_gamma_1}, we now explore the configuration where the MC is illuminated by an impulsive source of CRs, considering both 3D (Eq~\ref{eq:impulsive}) and 1D (Eq~\ref{eq:anisotropic_SNR}) diffusion.
The emission is still fully characterized by only three quantities: a normalization factor, the total spectral slope (respectively $\alpha+1.5 ~\delta$ and $\alpha+0.5 ~\delta$ for the 3D and 1D cases), and the energy-independent part of the exponential argument.
The latter parameter can be written as:

\begin{equation}
\label{eq:Lambda_gamma_imp_3}
     \Gamma_\gamma ^{imp}=\Gamma_{CR}^{imp} \left (  \frac{\rm GeV}{E}\right)^{\delta} =\frac{D_0}{10^{28}\,\rm cm^2/s}
~\frac{t}{10^3\,\rm yr}
~\left(\frac{50\,\rm pc}{R}\right)^2
\end{equation}
while the normalization factor is:

\begin{equation}
\label{eq:Lambda_gamma_imp}
    \Lambda_{\gamma,3D}^{imp}= \Lambda_{CR}^{imp}~\rm A_5
\end{equation}
\begin{equation}
\label{eq:Lambda_gamma_imp_2}
    \Lambda_{\gamma,1D}^{imp}= \Lambda_{CR,~1D}^{imp}~\rm A_5
\end{equation}
where $A_5=(M/10^5M_{\odot}) \left({\rm kpc}/d\right)^2$ represents the MC mass-to-distance-squared ratio, normalized to reference values.
Since the effective spectral slopes differ between the 3D and 1D cases, we fix $\delta=0.3$ when comparing the two models and vary only $\alpha$ instead. We also impose $R_d < L \sim 100~\rm pc$, beyond which Eq.~\ref{eq:anisotropic_SNR} is not longer valid. 

\begin{figure*}
   \centering
   \includegraphics[width=0.99\textwidth]{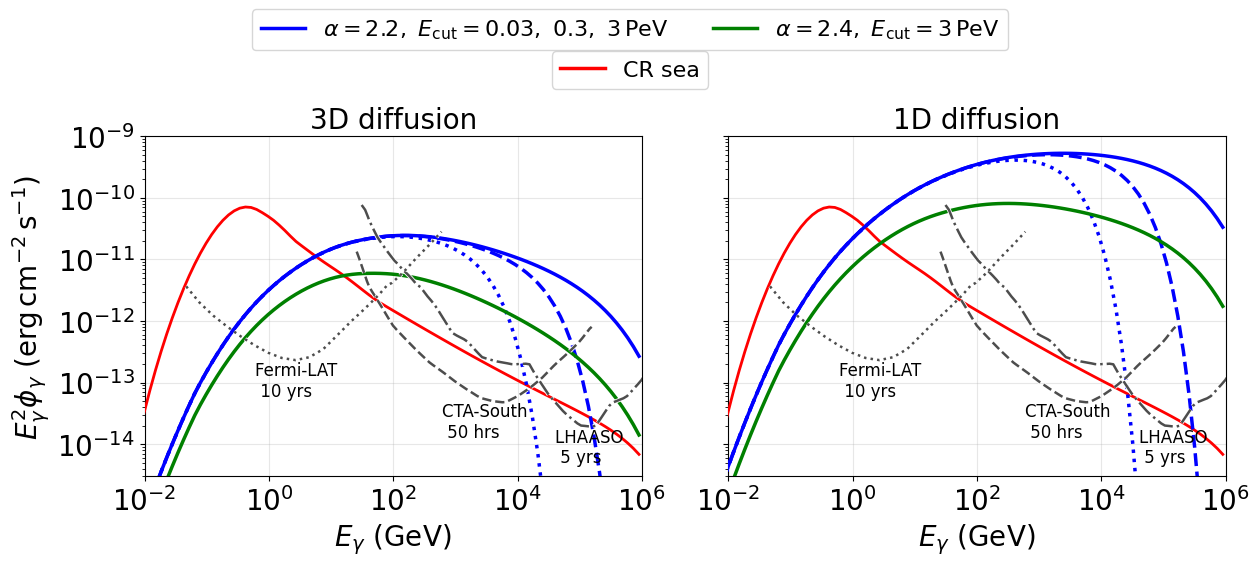}
   \caption{Same than Fig~\ref{fig:gamma_wind_steady} but for an impulsive source (considering $A_5=\Lambda_{CR}^{imp} = \Lambda_{CR,1D}^{imp} =\Gamma_{\gamma}^{imp}=1$) and varying only the injection slope $\alpha$ and the exponential cutoff energy.}
    \label{fig:gamma_snr_fixed_t}%
 \end{figure*}

Fig.~\ref{fig:gamma_snr_fixed_t} shows the gamma-ray emission for a benchmark system with $A_5 = 1$ and $\Lambda_{\gamma,3D} = 1$ (left panel) or $\Lambda_{\gamma,1D} = 1$ (right panel). An exponential cutoff is introduced in the CR proton spectrum to account for the finite maximum energy attainable by the accelerator. The dotted, dashed, and solid lines correspond to maximum energies of 0.03, 0.3, and 3 PeV, respectively.

The blue and green lines show the gamma-ray emission from the cloud produced by interactions of CRs that have escaped the accelerator, for two different injection slopes ($\alpha = 2.2$ and 2.4). 
The total gamma-ray emission from the cloud also includes the contribution from the Galactic CR sea, shown by the red line.

\begin{figure*}
   \centering
   \includegraphics[width=0.99\textwidth]{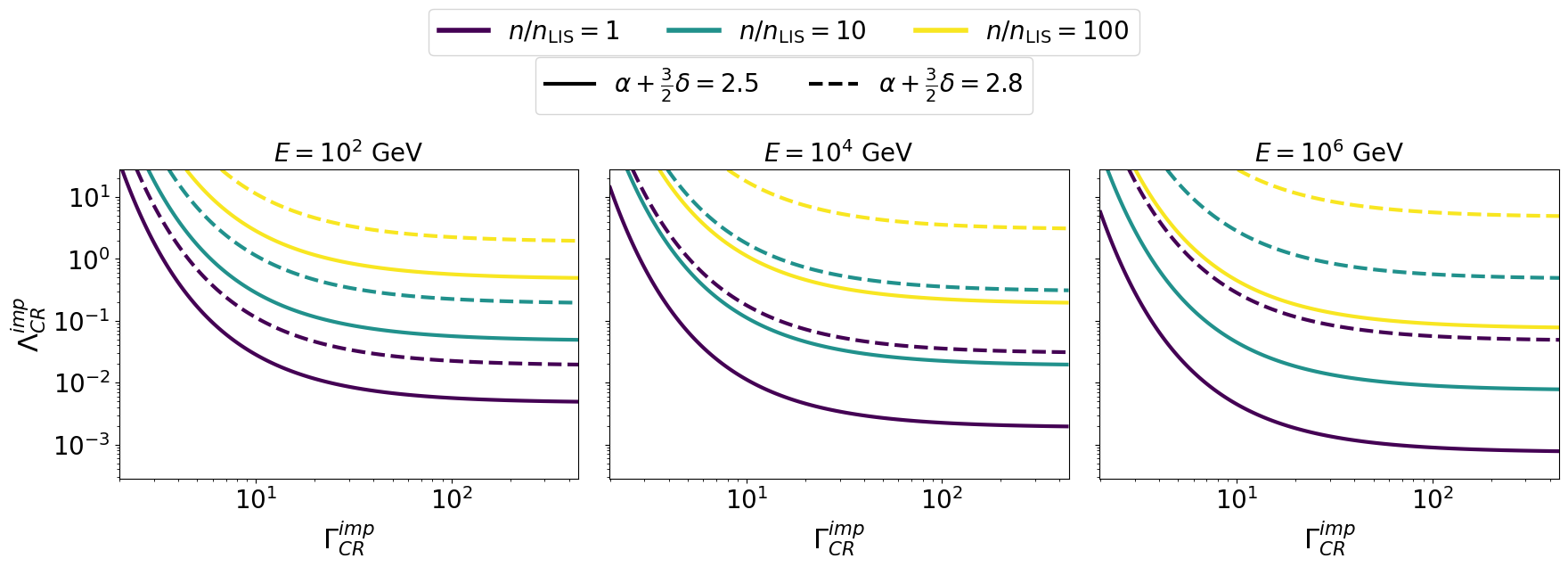}
   \caption{Same than Fig.~\ref{fig:adim_par_wind} but for the impulsive case, using the values of $\Lambda_{CR}^{imp}$, $\Gamma_{CR}^{imp}$ (Eq.~\ref{eq:Lambda_CR_imp}, \ref{eq:Gamma_CR_imp}) and $\alpha+1.5~\delta$.}
    \label{fig:adim_par_snr}%
 \end{figure*}

As anticipated from Fig.~\ref{fig:adim_par_snr}, for isotropic diffusion the CR sea contribution dominates over the source one only at low energies.
The source contribution largely exceeds the CR sea at higher energies ($E_{\gamma} \gtrsim 10~\mathrm{GeV}$) in the benchmark cases, but would drop and shift towards the left as time increases.
For anisotropic diffusion, the spectral shape remains qualitatively similar, but the flux is significantly enhanced. Consequently, the emission remains well above the detection thresholds of the considered instruments in all these benchmark cases.

\subsection{Complete gamma-ray characterization}

Even more in the impulsive case than in the continuous one, characterizing the temporal evolution of the system is essential to complement the benchmark scenario presented in Fig.~\ref{fig:gamma_snr_fixed_t}.

\begin{figure}
   \centering
   \includegraphics[width=0.45\textwidth]{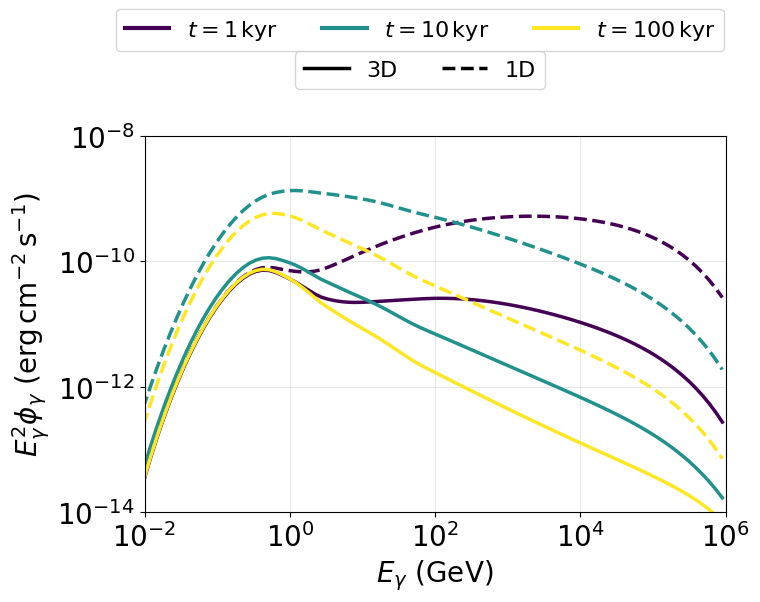}
   \caption{Same than Fig.~\ref{fig:gamma_wind_time} but with impulsive injection.
   Parameters are fixed to $W_{CR}\rm A_5= 10^{50} \rm ~erg$, $\rm R=50~pc$, $L= 100~\rm pc$, $D_0=10^{28} \rm cm^2/s$, $\delta=0.3$, $\alpha=2.2$.}
    \label{fig:gamma_snr_time}%
 \end{figure}

In this regard, Fig.~\ref{fig:gamma_snr_time} shows the temporal evolution of the gamma-ray spectra obtained by summing the contributions from escaped CRs and the CR sea, for 3D diffusion (Eq.~\ref{eq:impulsive}) and 1D diffusion (Eq.~\ref{eq:1D_snr_time}), represented respectively by solid and dashed curves. Source ages ranging from $10^3$ yr to $10^5$ yr are considered, with all other parameters fixed as indicated in the figure caption. As it can be seen from this figure, the spectral peak progressively shifts towards lower energies as time increases, while its amplitude decreases after a large time.

\begin{figure} 
    \centering
    \includegraphics[width=0.95\columnwidth]{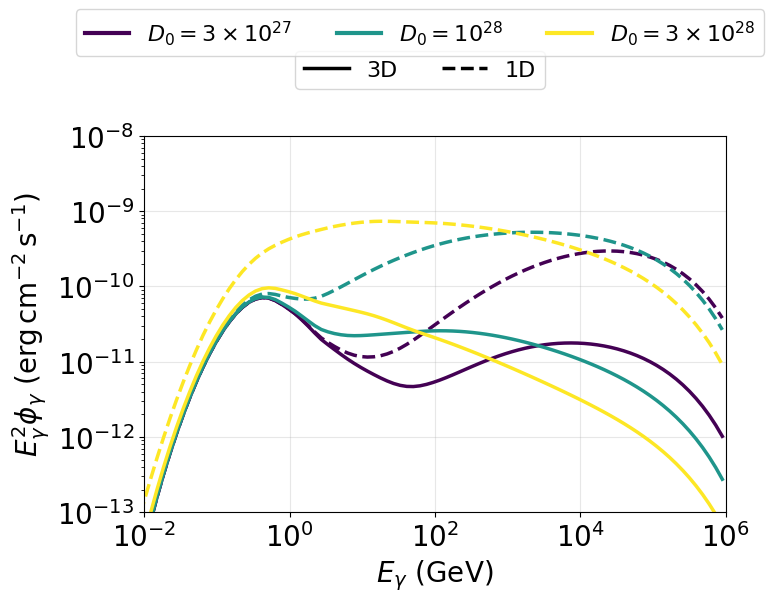}
    \caption{Same than Fig.~\ref{fig:gamma_snr_time} but at a fixed time $t=10^3 ~\rm yrs$ and with different colors representing different diffusion coefficient normalization (in units of $\rm cm^2/s$).}
    \label{fig:flux_snr_gamma_D0}
\end{figure}

Beyond this temporal evolution, a complete characterization requires independently varying the other parameters entering $\Gamma_{CR}^{imp}$.
First, Fig.~\ref{fig:flux_snr_gamma_D0} investigates the impact of varying the diffusion coefficient normalization $D_0$ (from $3\times 10^{27}$ to $3\times 10^{28}$), while keeping all other parameters fixed. Since the dependence on $D_0$ is degenerate with that on time, the resulting behavior is qualitatively similar to the temporal evolution discussed above. 

\begin{figure} 
    \centering
    \includegraphics[width=0.95\columnwidth]{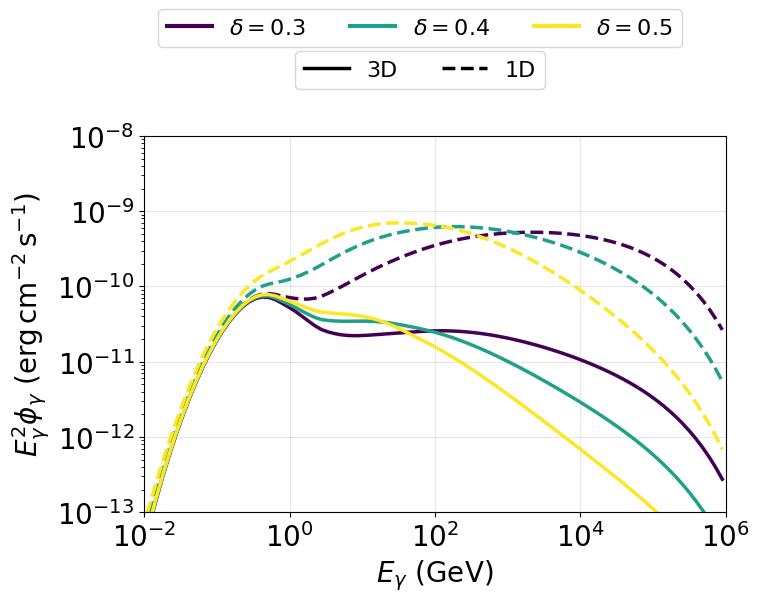}
    \caption{Same than Fig.~\ref{fig:gamma_snr_time} but at a fixed time $t=10^3 ~\rm yrs$ and with different colors representing different diffusion coefficient energy slopes.}
    \label{fig:flux_snr_gamma_delta}
\end{figure}

\begin{figure}
    \centering
    \includegraphics[width=0.95\columnwidth]{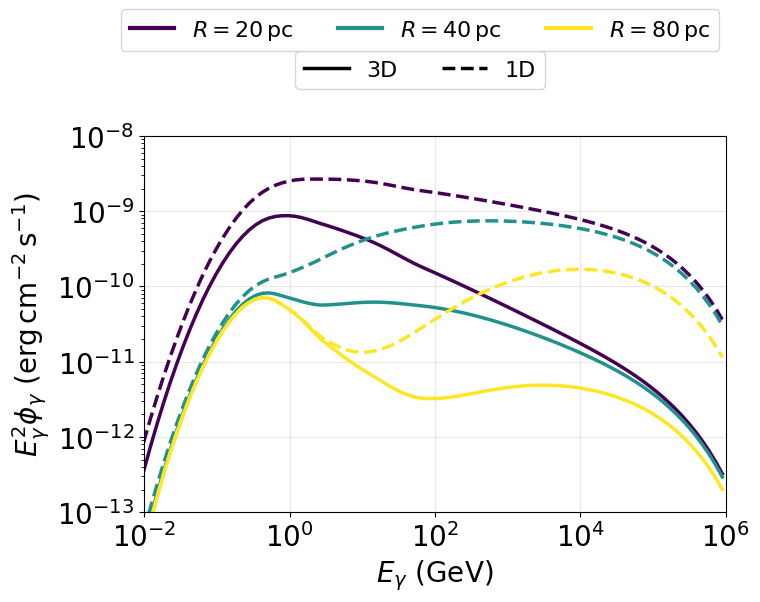}
    \caption{Same than Fig.~\ref{fig:gamma_snr_time} but at a fixed time $t=10^3 ~\rm yrs$, with different colors representing different source-cloud distance.}
    \label{fig:flux_snr_gamma_radius}
\end{figure}

Fig.~\ref{fig:flux_snr_gamma_delta} examines the impact of varying the diffusion index $\delta$, and Fig.~\ref{fig:flux_snr_gamma_radius} shows the effect of varying the source-cloud distance $R$. Since $R$ enters only through $\Gamma_{CR}^{imp}$ and does not affect the normalization factor, the resulting behavior is simpler: increasing the distance leads to an increasingly strong suppression of the emission at low energies.

\clearpage

\section{X-ray analysis}
\label{Appendix:X-rays}

\begin{figure}[h!]
    \centering
    \includegraphics[width=0.95\columnwidth]{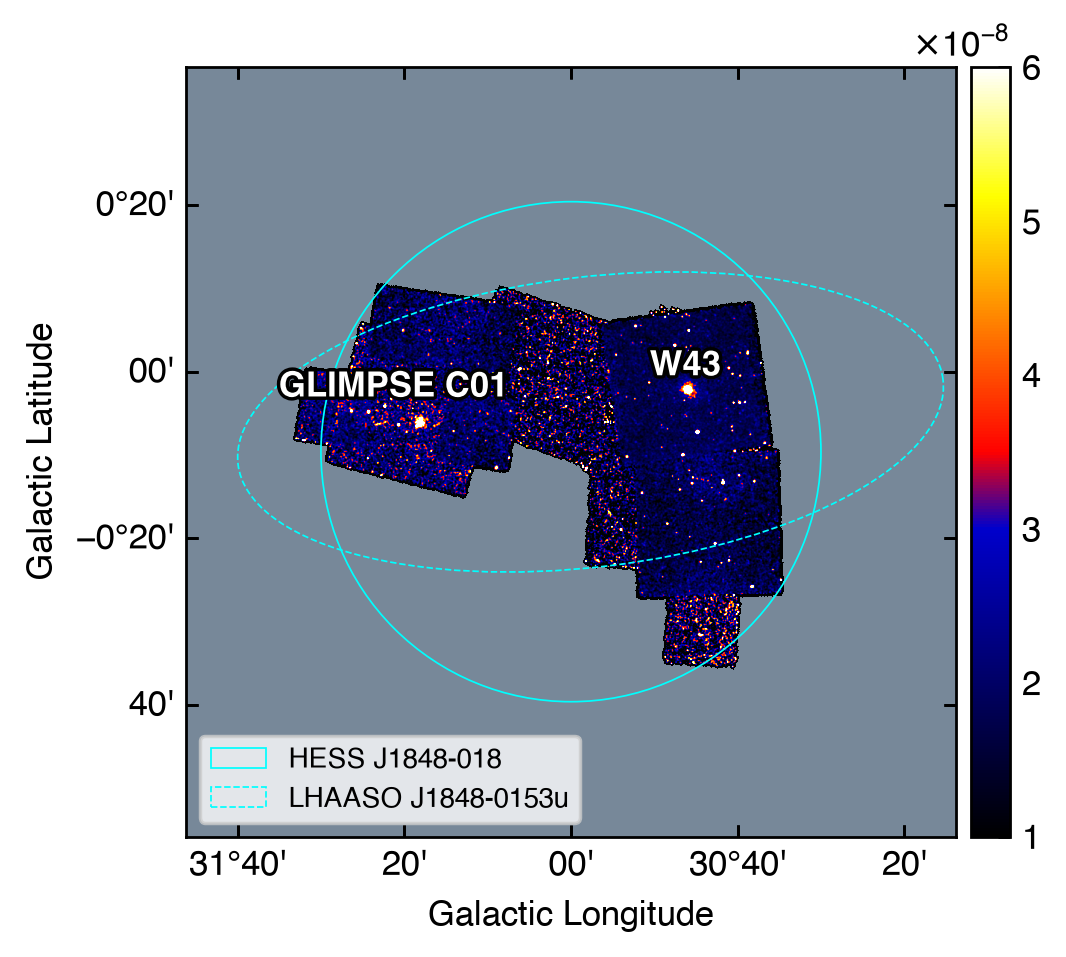}
    \caption{
    The Chandra flux image in 0.5--7 keV in units of counts~cm$^{-2}$~s$^{-1}$. One pixel corresponds to 2\arcsec.
    The gamma-ray emitting regions with H.E.S.S. and LHAASO (KM2A) are indicated by the solid circle and dashed ellipse, respectively.
    }
    \label{fig:chandra}
\end{figure}

The X-ray flux shown in Fig.~\ref{fig:w43_gamma_IC} was derived as follows.
We made use of archival Chandra data; ObsIDs 17716, 17717, 18867, 18868, 18869, 18870, 18887, and 18888 pointing at W43 (345 ks in total),
6587, 21641, and 15359 at HESS~J1848$-$018 North (86 ks),
and 12152, 17718, and 18890 at HESS~J1848$-$018 South east (108 ks).
The data were processed using CALDB version 4.9.6 in CIAO version 4.14.
All observation epochs were combined by using {\tt merge\_obs}.
Fig.~\ref{fig:chandra} shows a Chandra flux image in 0.5--7 keV, produced by {\tt fluximage}.
Almost half of the gamma-ray emission region of HESS~J1848$-$018 has been covered by Chandra, while the rest remains unexplored in the archival X-ray data (including XMM-Newton).
Point sources were identified by using {\tt wavedetect} and excluded in the following spectral analysis.
There are no apparent X-ray diffuse emission.
From each pointing (W43, HESS~J1848$-$018 North and SE), we extracted spectra and fit with a source model and a background model. The source model is described by an absorbed \cite[TBabs model; ][]{Wilm2000} power law.
The background model consists of instrumental background \citep{Bartalucci2014}, Cosmic X-ray background, foreground emission, and Galactic Ridge X-ray emission \citep{Uchiyama2013}.
With a column density $N_{\rm H}$ and a photon index $\Gamma$ being fixed to (2--4)$\times 10^{22}~{\rm cm}^{-2}$ \citep{Arora2020} and 2, respectively, we obtained 1--10 keV flux upper limits of $\sim 5 \times 10^{-16}$~erg~cm$^{-2}$~s$^{-1}$~arcmin$^{-2}$. 
This yields $\sim 1 \times 10^{-12}$~erg~cm$^{-2}$~s$^{-1}$ in Fig.~\ref{fig:w43_gamma_IC} by scaling to $r=0.5$~deg \citep{HESS}.

\end{appendix}

\end{document}